\documentclass[12pt] {article}
\usepackage{epsfig,graphicx,psfrag}

\normalsize
\newcommand{\gsim}{\hbox{ \raise3pt\hbox to 0pt{$>$}\raise-3pt\hbox{$\sim $} }}
\newcommand{\simgt}{\hbox{ \raise3pt\hbox to 0pt{$>$}\raise-3pt\hbox{$\sim$} }}
\newcommand{\lsim}{\hbox{ \raise3pt\hbox to 0pt{$<$}\raise-3pt\hbox{$\sim $} }}
\newcommand{\simlt}{\hbox{ \raise3pt\hbox to 0pt{$<$}\raise-3pt\hbox{$\sim$} }}
\newcommand{\be}{\begin{equation}}
\newcommand{\ee}{\end{equation}}
\newcommand{\bea}{\begin{eqnarray}}
\newcommand{\eea}{\end{eqnarray}}
\newcommand{\mathbold}[1]{\mbox{\boldmath $\bf#1$}}

\def\to{\rightarrow}
\newcommand{\figdir}{figs}
\begin{document}
%
\begin{titlepage}

\hfill
\begin{minipage}[t]{4.7cm}
\begin{flushleft}
{\bf KEK Preprint 2002-11}\\
{\bf HUBP-01/02}\\
{\bf TU-651}\\
{\bf April 2002}\\
\end{flushleft}
\end{minipage}

\vspace*{1.0cm}

\begin{center}
{\Large \bf
Kinematical Reconstruction of the $\mathbold{\it{t\bar{t}}}$ System\\
Near its Threshold at Future $\mathbold{\it{{e^+e^-}}}$ Linear Colliders
}
\end{center}

\vskip 1.5cm

\begin{center}
Katsumasa Ikematsu$^{a)}$
\footnote[1]{e-mail address: \texttt{ikematsu@post.kek.jp}},
Keisuke Fujii$^{b)}$
\footnote[2]{e-mail address: \texttt{fujiik@jlcuxf.kek.jp}},
Zenr\=o Hioki$^{c)}$
\footnote[3]{e-mail address: \texttt{hioki@ias.tokushima-u.ac.jp}},\\
Yukinari Sumino$^{d)}$
\footnote[4]{e-mail address: \texttt{sumino@tuhep.phys.tohoku.ac.jp}},
Tohru Takahashi$^{e)}$
\footnote[5]{e-mail address: \texttt{tohrut@hiroshima-u.ac.jp}}
\end{center}

\vskip 0.7cm

\begin{center}
a) Department of Physics, Hiroshima University,\\
Higashi Hiroshima 739-8526, Japan
\end{center}
\begin{center}
b) IPNS, KEK, Tsukuba 305-0801, Japan
\end{center}
\begin{center}
c) Institute of Theoretical Physics, University of Tokushima,\\
Tokushima 770-8502, Japan
\end{center}
\begin{center}
d) Department of Physics, Tohoku University,\\
Sendai 980-8578, Japan
\end{center}
\begin{center}
e) Graduate School of Advanced Sciences of Matter, Hiroshima University,\\
Higashi Hiroshima 739-8530, Japan
\end{center}

\vskip 1.5cm
\centerline{\bf ABSTRACT}
\vskip 0.7cm
\normalsize
We developed a new method for full kinematical
reconstruction of the $t\bar{t}$ system near its threshold
at future linear $e^+e^-$ colliders.
In the core of the method lies likelihood
fitting which is designed to improve
measurement accuracies of the kinematical
variables that specify the final states
resulting from $t\bar{t}$ decays.
The improvement is demonstrated by applying
this method to a Monte-Carlo
$t\bar{t}$ sample generated with
various experimental effects including beamstrahlung,
finite acceptance and resolution of the detector
system, etc.
A possible application of this method
and its expected impact are also discussed.
\end{titlepage}

%
\section{Introduction}

The discovery of the top quark~\cite{CDF-D0:1994} at Tevatron
has completed the standard-model (SM) list of matter fermions.
In spite of its subsequent studies thereat, however,
our knowledge on its properties is still far below the level
we reached for the lighter matter fermions.
A next-generation $e^+e^-$ linear collider such as JLC~\cite{JLC1},
having a facet as a top-quark factory,
is expected to allow us to measure
top quark's properties with unprecedented precision,
thereby improving this situation dramatically.
Such precision measurements may shed light on
the electroweak-symmetry-breaking
mechanism or hint beyond-the-SM physics or both.

Being aware of the opportunities provided by the linear collider,
a number of authors have so far taken up this subject and performed
interesting analyses~\cite{Atwood:2000tu,ACFARep,
Aguilar-Saavedra:2001rg,Abe:2001wn}.
These analyses can be classified into two categories, i.e.,
those near the $t\bar{t}$ threshold, mainly focused
on physics contained in the threshold enhancement factor,
and those in open-top region,
searching for anomalies in production and decay vertices,
both of which play important roles
and complement each other.
In order to thoroughly carry out such analyses for real data,
we need a sophisticated method to kinematically reconstruct
events as efficiently and as precisely as possible.
This is, however, highly non-trivial in practice,
due to finite detector resolutions,
possible missing neutrinos in the final states,
and various background contributions.
We thus need to further explore the potential
of the $e^+e^-$ linear collider and extend
the past studies~\cite{Fujii:1993mk,Barklow:1995,Frey:1997sg}
to make maximum use
of the linear collider's advantages:
clean experimental environment, well-defined initial state,
availability of highly polarized electron beam,
possibility of full parton-level reconstruction of final states, etc.

As mentioned above, feasibility studies on form factor measurements
are mainly done in the open-top region
~\cite{Barklow:1995,Frey:1997sg,Iwasaki:2001ip}.
It has, however, been pointed out that
form factor measurements in the $t\bar{t}$ threshold region
have some favorable features such as availability of
well-controlled highly polarized top sample,
no need for transformation to $t$ or $\bar{t}$ rest frames
because both of the $t$ and $\bar{t}$ are
nearly at rest~\cite{Gusken:1985nf,Peter:1997rk}.
Besides, as far as the decay form factor measurements
of an on-shell top quark are concerned,
the center-of-mass energy does not matter.
Motivated by these observations,
we decided to study feasibility of form factor measurements at
the $t\bar{t}$ threshold.
From an experimental point of view,
the top quark physics at the $e^+e^-$ linear collider
is expected to commence near the $t\bar{t}$ threshold
and analysis techniques developed there are expected to
be easily extended to the open-top region.

In this paper, we thus aim at state-of-the-art analysis method such as a
new method to reconstruct kinematically the $t\bar{t}$ system near its
threshold in $e^+e^-$ annihilation.
The content will be divided into six
parts. In Sec.~2 we briefly review our simulation framework. Sec.~3 is
devoted to top quark reconstruction in the lepton-plus-4-jet mode, where
two subsections recapitulate basic strategy and procedure,
respectively. In Sec.~4 we explain our kinematical reconstruction
method which is designed to improve measurement accuracies of the
kinematical variables that specify the final states resulting from
$t\bar{t}$ decays. Then we discuss a possible application of this method
and its expected impact in Sec.~5.
Finally, Sec.~6 summarizes our results and concludes this paper.

%
\section{Framework of Analysis}
\label{Sec:framework}

For Monte-Carlo-simulation studies of $t\bar{t}$ productions and decays,
we developed an event generator
that is now included in
\texttt{physsim-2001a}~\cite{PHYSSIM},
where the amplitude calculation and phase space integration are
performed with \texttt{HELAS/BASES}~\cite{Murayama:1992gi,Kawabata:1985yt}
and parton 4-momenta of an event are generated by
\texttt{SPRING}~\cite{Kawabata:1985yt}.
In the amplitude calculation,
initial state radiation (ISR) as well as
$S$- and $P$-wave QCD corrections to the $t\bar{t}$ system
~\cite{Sumino:1992ai,Murayama:1992mg}
are taken into account.
Parton showering and hadronization are carried out
using \texttt{JETSET 7.4}~\cite{Sjostrand:1993yb}
with final-state tau leptons treated by \texttt{TAUOLA}~\cite{Jadach:1993hs}
in order to handle their polarizations properly.

In this study,
the top-quark (pole) mass is assumed to be 175~GeV
and the nominal center-of-mass energy is set at
2~GeV-above the $1S$ resonance of the $t\bar{t}$ bound states.
This energy is known to be
suitable for measurements of various properties
of the $t\bar{t}$ system at threshold~\cite{Fujii:1993mk}.
We will assume an electron-beam polarization of 80\%
in what follows.
Effects of natural beam-energy spread and beamstrahlung
are taken into account according to the prescription given in
\cite{Fujii:1993mk}, where
the details of the beam parameters are also described.
We have assumed no crossing angle between
the electron and the positron beams
and ignored the transverse component of the initial state radiation.
Consequently,
the $t\bar{t}$ system in our Monte-Carlo sample
has no transverse momentum.
Under these conditions we expect 40k
$t\bar{t}$ events for 100$fb^{-1}$.

The generated Monte-Carlo $t\bar{t}$ events were passed to a detector
simulator (\texttt{JSF Quick Simulator}~\cite{JSF})
which incorporates the ACFA-JLC study parameters
(see Table.~\ref{det-param}).
The quick simulator created vertex-detector hits,
smeared charged-track parameters in the central tracker with
parameter correlation properly taken into account,
and simulated calorimeter signals as from individual segments,
thereby allowing realistic simulation of cluster overlapping.
It should also be noted that track-cluster matching was
performed to achieve the best energy-flow measurements.

\begin{table}[h]
 \begin{center}
   \begin{tabular}{|c||r|r|} \hline
     \textbf{Detector} & \textbf{Performance} & \textbf{Coverage} \\
     \hline
     \hline
     Vertex detector                                                      &
       $ \sigma_{\mathrm{b}}
         = 7.0 \oplus (20.0/p) \, / \, \sin^{3/2}\theta \; \mu\mathrm{m}$ &
       $ | \cos\theta | \leq 0.90 $ \\
     \hline
     Central drift chamber                                                &
       $ \sigma_{p_{T}}/p_{T}
         = 1.1 \times 10^{-4} p_{T} \> \oplus \> 0.1 \; \% $              &
       $ | \cos\theta | \leq 0.95 $ \\
     \hline
     EM calorimeter                                                       &
       $ \sigma_{E}/E
         = 15 \; \% \> / \sqrt{E} \> \oplus 1 \; \% $                     &
       $ | \cos\theta | \leq 0.90 $ \\
     \hline
     Hadron calorimeter                                                   &
       $ \sigma_{E}/E
         = 40 \; \% \> / \sqrt{E} \> \oplus 2 \; \% $                     &
       $ | \cos\theta | \leq 0.90 $ \\
     \hline
   \end{tabular}
 \end{center}
 \caption[]{
       ACFA study parameters of the JLC detector, where
       $p$, $p_{T}$, and $E$ are measured in units of GeV.
       }
 \label{det-param}
\end{table}%

%
\section{Event Selection}

\subsection{Basic Reconstruction Strategy}

Since the top quark decays almost 100\% into
a $b$ quark and a $W$ boson, the signature of
a $t\bar{t}$ production is two $b$ jets and two
$W$ bosons in the final state.
These $W$ bosons decay subsequently either
leptonically into a lepton plus a neutrino
or hadronically into two jets.
According to how the $W$ bosons decay, therefore,
there will be three modes of final states:
(1) six jets, where both of the $W$'s decay hadronically,
(2) one lepton plus four jets, where one of the $W$'s decays
leptonically and the other hadronically, and
(3) two leptons plus two jets, where both of the $W$'s decay
leptonically.

In order to reconstruct the momentum vector of the top quark,
we will use the lepton-plus-4-jet mode,
for which we can reconstruct the $t$($\bar{t}$)-quark momentum
as the momentum sum of the $b$($\bar{b}$) jet and
the two jets from the hadronically-decayed $W^+$($W^-$),
while we can tell the charge of the hadronically-decayed
$W$ from the charge of the lepton.

In the lepton-plus-4-jet mode,
two of the four jets are $b$($\bar{b}$) jets directly
from the $t$($\bar{t}$) quarks, while the other two are from
the $W$ boson that decayed hadronically.
Therefore, if one can identify the $b$ and $\bar{b}$ jets,
remaining two jets can be uniquely assigned
as decay products of the $W$ boson.
The other $W$ boson can be reconstructed from
the lepton and the neutrino indirectly detected
as a missing momentum.
Remaining task is then to decide which $b$($\bar{b}$) jet
to attach to which $W$-boson candidate,
in order to form $t$($\bar t$) quarks.
Since the $t$($\bar t$) quarks are virtually at rest near the threshold,
a $b$($\bar b$) jet and the corresponding $W$ boson fly
in the opposite directions.
We can thus choose the correct combination by
requiring the $b$($\bar b$) jet and the $W$ boson
be approximately back-to-back.

In reality, however,
$b$($\bar b$)-quark tagging is not perfect and
can be performed only with some finite efficiency and purity:
there could be more than two $b$($\bar{b}$)-jet candidates
in a single event.
In addition, $b$ and $\bar{b}$ quarks can be
emitted in the same direction.
In such a case, a wrong combination could
accidentally satisfy the back-to-back condition.
These facts sometimes prevent us from uniquely assigning each jet to
its corresponding parton, resulting in multiple solutions
for a single event.
Moreover, the leptonically-decayed $W$ is poorly
reconstructed in practice,
since the neutrino momentum is strongly affected
by ISR, beamstrahlung, as well as other possible
neutrinos emitted from the $b$($\bar{b}$) jets.
In order to overcome these difficulties,
we will need some sophisticated method.
We defer discussion of such a method to the next section and
examine here the extent to which
the aforementioned basic reconstruction strategy works.

\subsection{Event Selection Procedure}

\begin{figure}[hp]
 \begin{center}
   \includegraphics[height=6cm,clip]{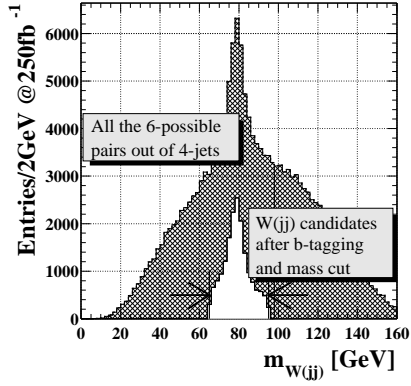}
 \end{center}
 \caption{Invariant mass distributions of the 2-jet systems
          reconstructed as $W$-boson candidates.
          Hatched and solid histograms correspond to before and after
          double $b$-tagging, respectively.
	  The locations of the $W$ mass cuts
	  are indicated with arrows.}
 \label{M2j}
\end{figure}

The lepton-plus-4-jet-mode selection started with demanding
an energetic isolated lepton:
$E_\ell > 18~{\rm GeV}$ and $E_{14^\circ {\rm cone}} < 18~{\rm GeV}$,
where $E_\ell$ is the lepton's energy
and $E_{14^\circ {\rm cone}}$ is the energy sum of particles
within a cone with a half angle of $14^\circ$ around the lepton direction
excluding the lepton itself.\footnote{
The $E_\ell$ cut was chosen to be the kinematical limit for the lepton
from the $W \to \ell\nu$ decay. On the other hand, the cone-energy cut was
optimized to achieve high purity, while keeping reasonable efficiency.}
When such a lepton was found,
the rest of the final-state particles was forced
clustering to four jets,
using the Durham clustering algorithm~\cite{Catani:1991hj}.
Two-jet invariant mass was then calculated for
each of the six possible combinations
and checked if it was between 65~GeV and 95~GeV,
in order to select a jet pair which was consistent
with that coming from a $W$-boson decay.
For such a jet pair the remaining two jets, at the same time,
had to be identified as $b$($\bar{b}$) jets,
using flavor tagging based on the impact parameter method.
The hatched histogram in Fig.~\ref{M2j} is the 2-jet invariant mass
distribution of all the possible pairs out of the four jets,
while the solid histogram being that with the $b$-tagging.
It is seen that this procedure dramatically improved the
purity of the $W$ boson sample.
It should also be stressed that these selection
criteria are very effective to suppress background processes
such as $e^{+}e^{-} \to W^{+}W^{-}$ and
provide us with an essentially background-free
$t\bar{t}$ event sample.

\begin{figure}[hp]
 \begin{center}
   \includegraphics[height=6cm,clip]{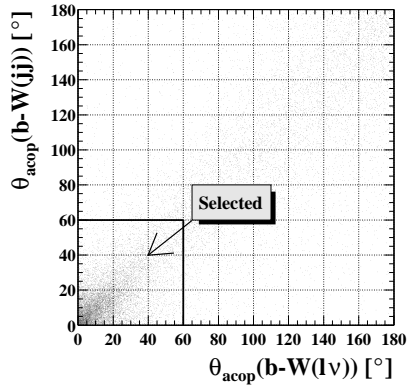}
 \end{center}
 \caption{Scatter plot of the acoplanarity angles
          corresponding to two $b$-$W$ systems, where horizontal and
          vertical axes are angles
          of $b$-$W_{\ell\nu}$ and $b$-$W_{2-jet}$ systems,
          respectively.}
 \label{AcopbW}
\end{figure}

The remaining task is to decide which $b$($\bar{b}$) jet
to associate with which $W$ candidate.
For a $b$($\bar{b}$)-jet candidate,
the right $W$ boson partner was selected
by requiring the back-to-back condition as described above.
Fig.~\ref{AcopbW} is a scatter plot of the acoplanarity angles
of the two possible $b$-$W$ systems
where horizontal and vertical axes are the angles
of $b$-$W_{\ell\nu}$ and $b$-$W_{2-jet}$ system, respectively.
$b$-$W$ pairs having $\theta_{acop(b-W)} \leq 60^\circ$ was
regarded as daughters of the $t$($\bar t$) quarks.

The selection efficiency after all of these cuts
was found to be 15\% including the
branching fraction to the lepton-plus-4-jet mode of 29\%.

%
\section{Kinematical Fit}

The event selection described above yields a
very clean $t\bar{t}$ sample.
As noted above, however, the sample is
still subject to combinatorial backgrounds,
if we are to fully reconstruct the
final state by assigning each jet to a corresponding
decay daughter of the $t$ or $\bar{t}$ quark.
We thus need a well-defined criterion to
select the best from possible multiple solutions.
It is also desirable to improve the measurement
accuracies of those kinematical variables which
are suffering from effects of missing neutrinos
(such variables include momenta of $b$, $\bar{b}$
or the neutrino from a $W$ itself).

The $t\bar{t}$ system produced via $e^+e^-$ annihilation
is a heavily constrained system: there are many
mass constraints in addition to the usual 4-momentum
conservation.
At $e^+e^-$ linear colliders,
thanks to their well-defined initial state and
the clean environment,
we can make full use of these constraints and
perform a kinematical fit to select the best
solution and to improve the measurement accuracies of
the kinematical variables of the final-state partons.

\subsection{Parameters, Constraints, and Likelihood Function}

\begin{figure}[hp]
  \begin{center}
    \includegraphics[height=5cm,clip]{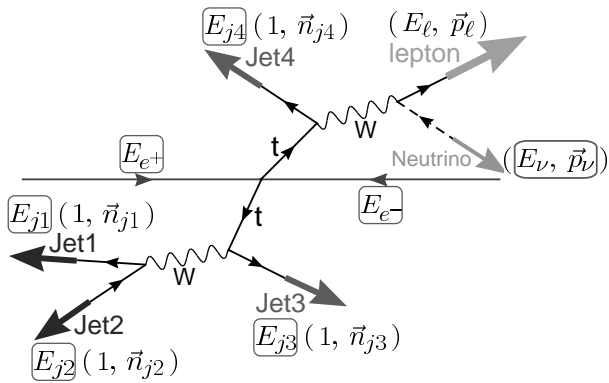}
  \end{center}
  \caption{
  	Schematic diagram showing parameters and constraints
  	relevant to the kinematical fit described in the text.
        The boxed parameters are unknown and
        to be determined by the fit.}
  \label{kine-fit}
\end{figure}

For the lepton-plus-4-jet final state,
there are 10 unknown parameters to be determined by the fit:
the energies of four jets,
the 4-momentum of the neutrino from the leptonically-decayed $W$ boson,
and the energies of the initial-state electron and positron,
provided that
the jet directions as output from the jet finder
are accurate enough,
the error in the 4-momentum measurement of the lepton
from the leptonically-decayed $W$ can be ignored,
and that the transverse momenta of the initial-state
electron and positron after beamstrahlung or initial-state
radiation or both are either negligible or known
from a low angle $e^+/e^-$ detector system\footnote{
In addition, there will be some finite transverse
momenta due to a finite crossing angle of the two beams.
These transverse momenta are, however, known and can
be easily incorporated into the fit.
}
(see Fig.~\ref{kine-fit}).

The requirements of 4-momentum conservation
and the massless constraint for the neutrino
from the leptonically-decayed $W$
reduce the number of free parameters to 5.
We choose, as these free parameters,
the energies of the four jets
and the initial longitudinal momentum
(the difference of the energies of
the initial-state electron and positron).

These five unknown parameters can be determined by
maximizing the following likelihood function:
\begin{equation}
  L = ( \prod_{f=1}^{4} P^f_{E_{f}}(E_{f}^{mean}, E_{f}))
      \cdot P_{\Gamma_{W^{+}}}
      \cdot P_{\Gamma_{W^{-}}}
      \cdot P_{\Gamma_{t\bar{t}}}
      \cdot P_{\sqrt{s}},
\end{equation}
where $P^f_{E_{f}}$ is a resolution function for jet $f$
and is Gaussian
for $f=1$ and $2$ (jets from the hadronically-decayed $W$)
as given by the detector energy resolution.
For $f=3$ and $4$ (jets from the $b$ and $\bar{b}$ quarks)
the resolution function is the same Gaussian convoluted
with the missing energy spectrum due to possible neutrino emissions.
For the two $W$ bosons in the final state,
we use a Breit-Wigner function $P_{\Gamma_{W}}$
instead of $\delta$-function-like mass constraints.
$P_{\sqrt{s}}$ is a weight function
coming from ISR and beamstrahlung effects.
This distribution was calculated as a differential cross section
as a function of the energies of initial-state electron and positron,
taking into account the $t\bar{t}$ threshold correction
as described in Sec.~\ref{Sec:framework}.

\begin{figure}[hp]
  \begin{center}
    \includegraphics[height=5cm,clip]{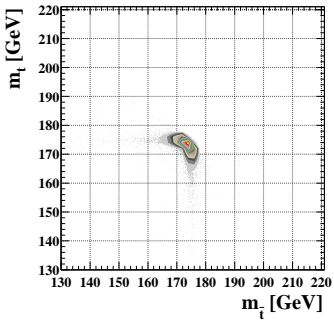}
  \end{center}
  \caption{$P_{\Gamma_{t\bar{t}}}$ distribution below $t\bar{t}$ threshold.}
  \label{Mttbar-gen}
\end{figure}

The remaining factor, $P_{\Gamma_{t\bar{t}}}$,
controls the mass distribution of the $t$ and $\bar{t}$ quarks
and has been introduced to take into account the
kinematical constraint that the $t$ and $\bar{t}$ cannot
be simultaneously on-shell below threshold
(see Fig.~\ref{Mttbar-gen} which shows $P_{\Gamma_{t\bar{t}}}$ distribution
below $t\bar{t}$ threshold).
$P_{\Gamma_{t\bar{t}}}$ distribution is a dynamics-independent factor
which is extracted from the theoretical formula
for the threshold cross section.

\subsection{Results}

We performed the maximum likelihood fit for the selected sample.
The maximum likelihood fit provided us with a well-defined
clear-cut criterion to select the best solution,
when there were multiple possible solutions for a single event:
we should select the one with the highest likelihood.

\begin{figure}[ht]
  \begin{center}
    \includegraphics[height=5cm,clip]{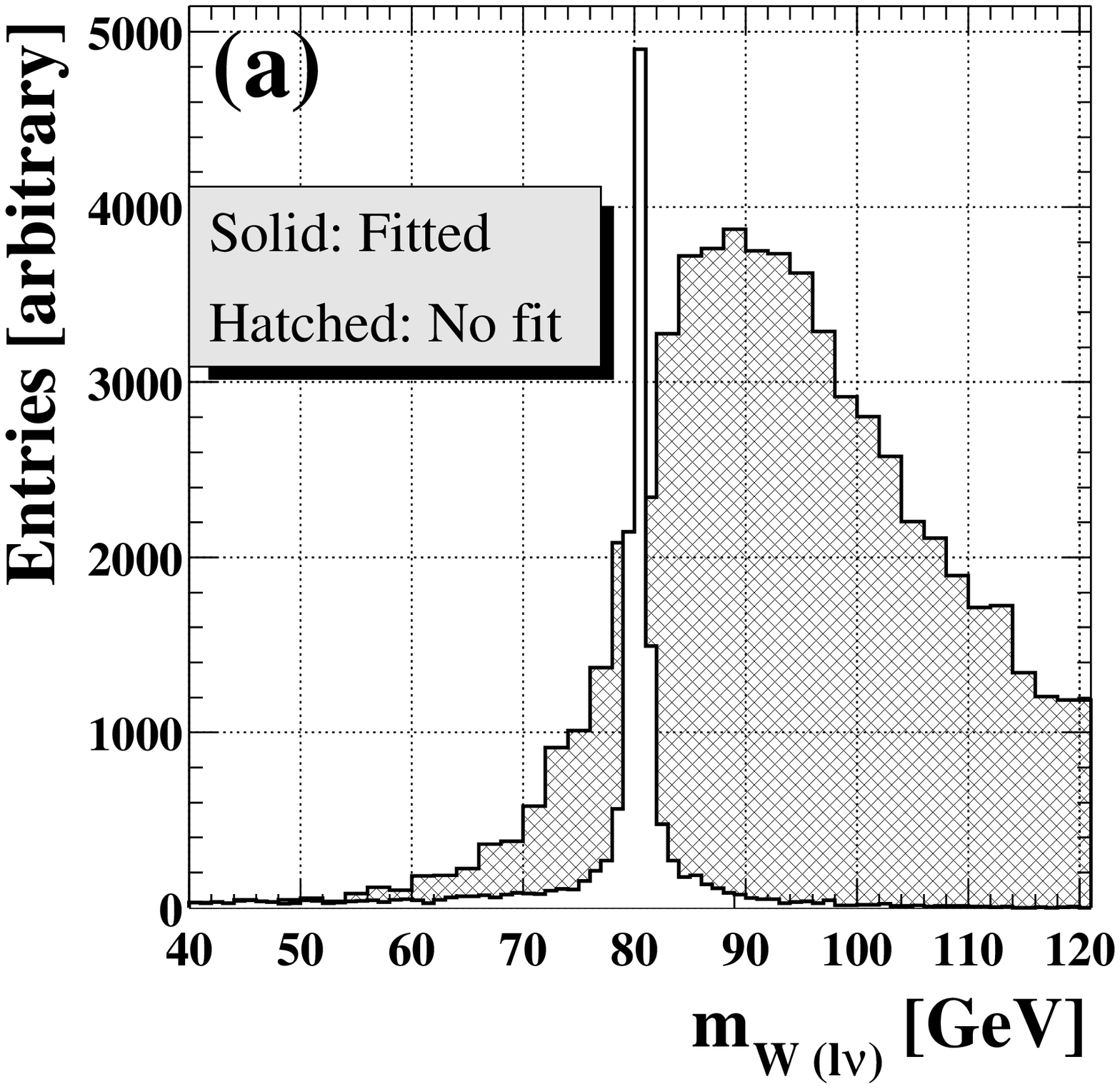}
    \includegraphics[height=5cm,clip]{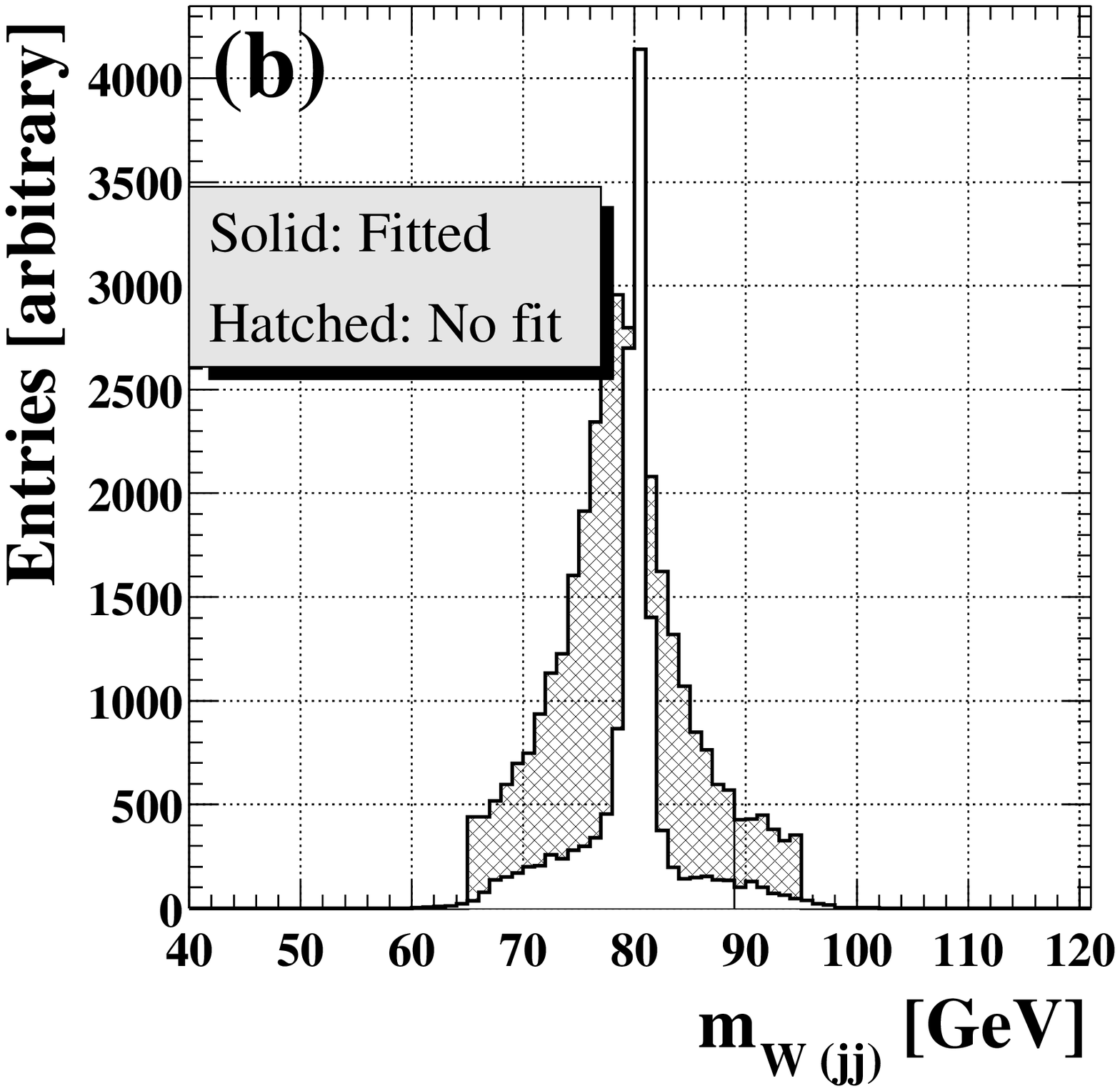}
  \end{center}
  \caption{
  	Reconstructed $W$ mass distributions for
  	(a) leptonically
  	and (b) hadronically-decayed $W$ bosons,
  	before (hatched) and after (solid)
  	the kinematical fitting.}
  \label{Mww}
\end{figure}

Figs.~\ref{Mww}-a) and -b) are the reconstructed $W$ mass
distributions for the leptonically and hadronically-decayed
$W$ bosons, respectively, before (hatched) and after (solid)
the kinematical fit.
The figures demonstrate that
the Breit-Wigner factors ($P_{\Gamma_{W^{\pm}}}$) in the
likelihood function properly constrain the $W$ masses
as intended.

\begin{figure}[hp]
  \begin{center}
    \includegraphics[height=5cm,clip]{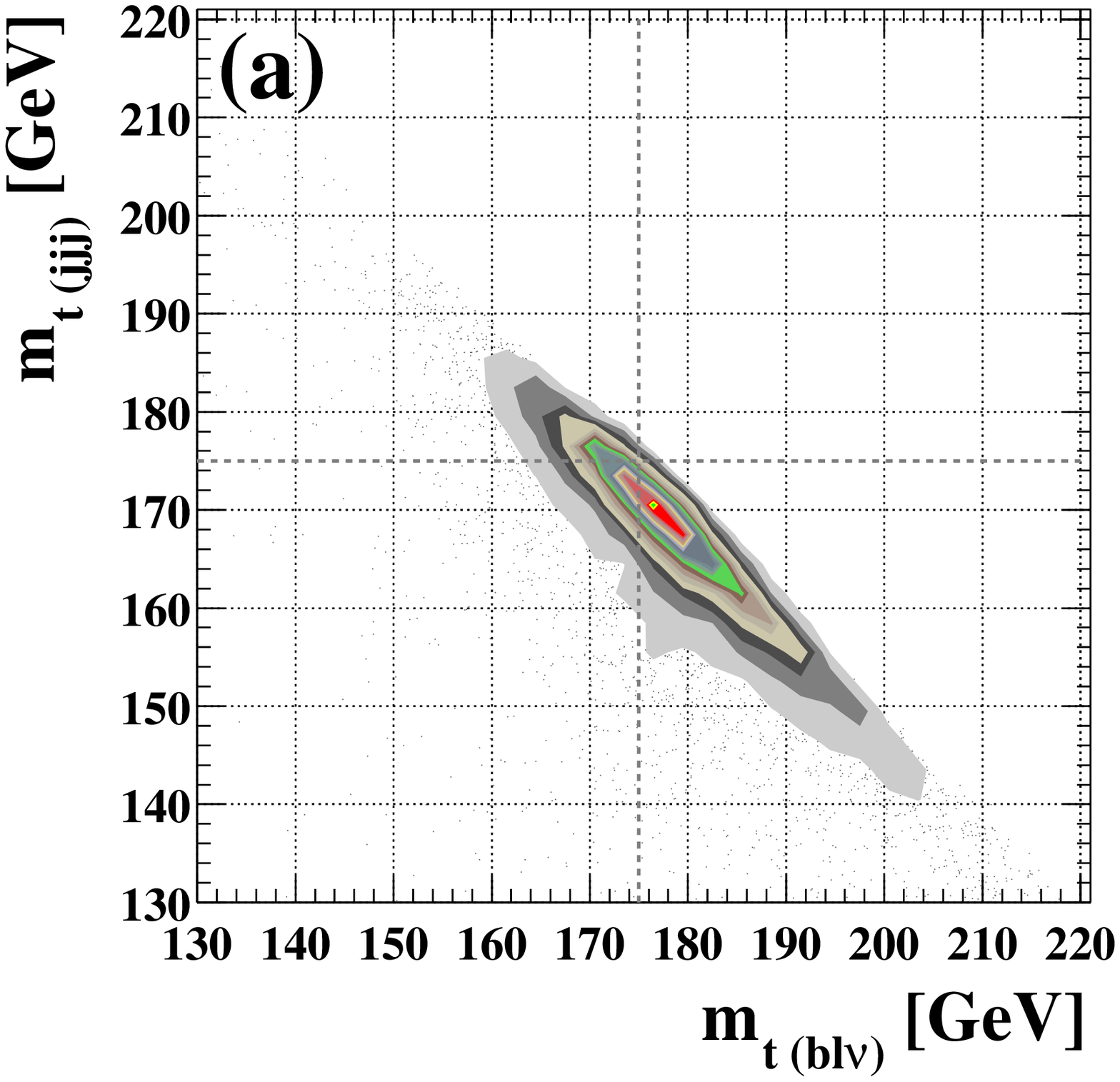}
    \includegraphics[height=5cm,clip]{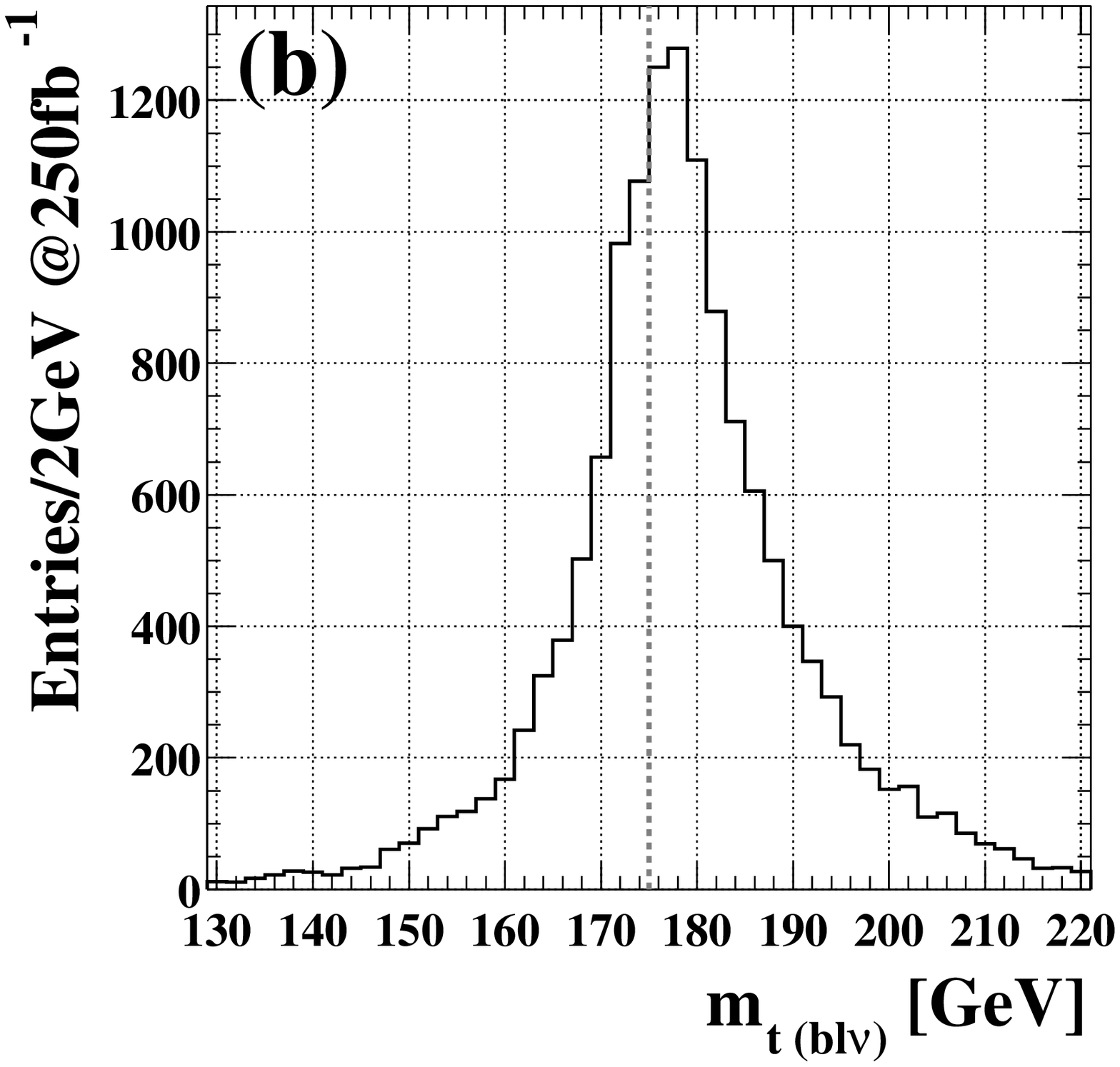}
    \includegraphics[height=5cm,clip]{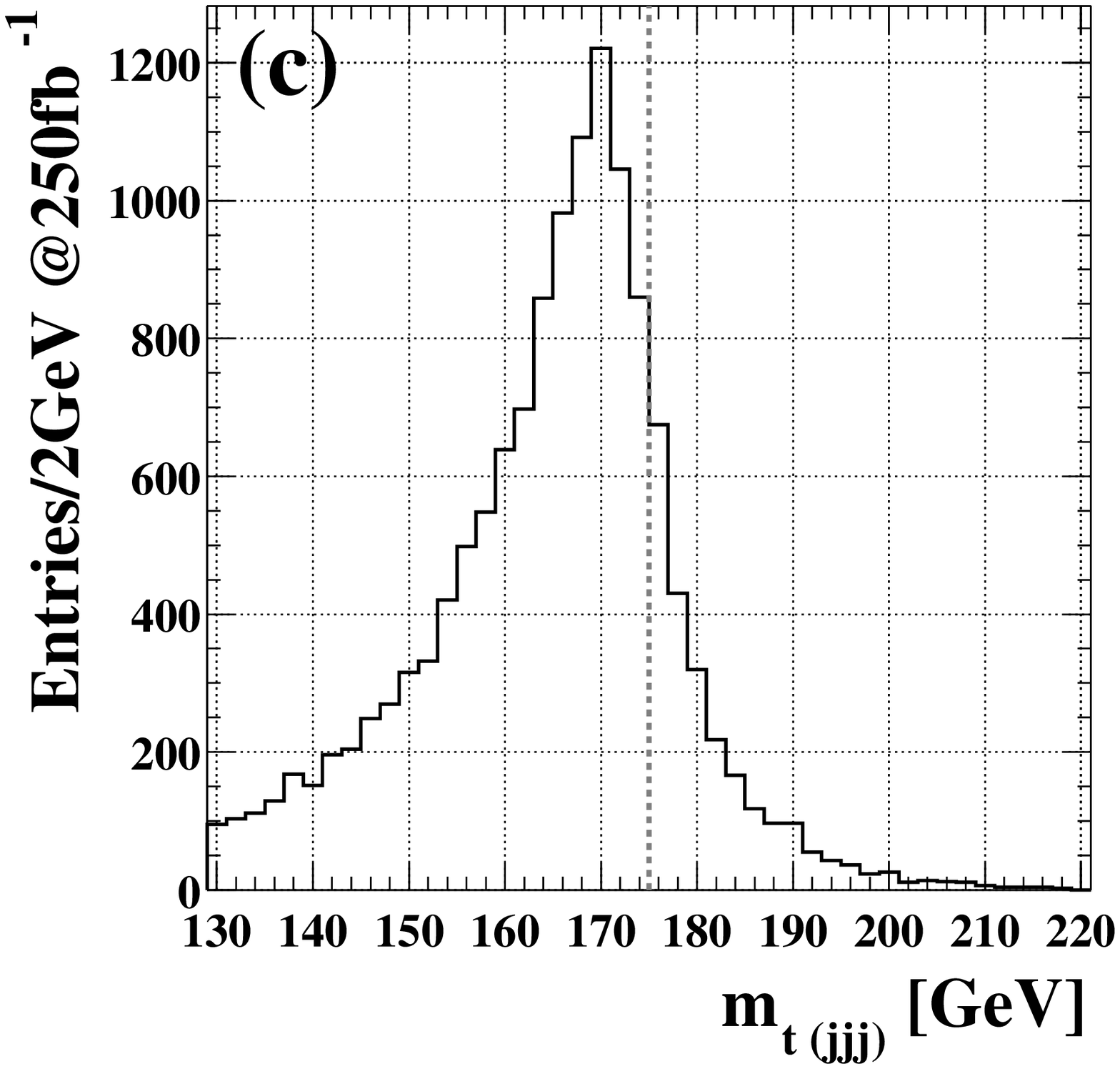}
    \\
    \includegraphics[height=5cm,clip]{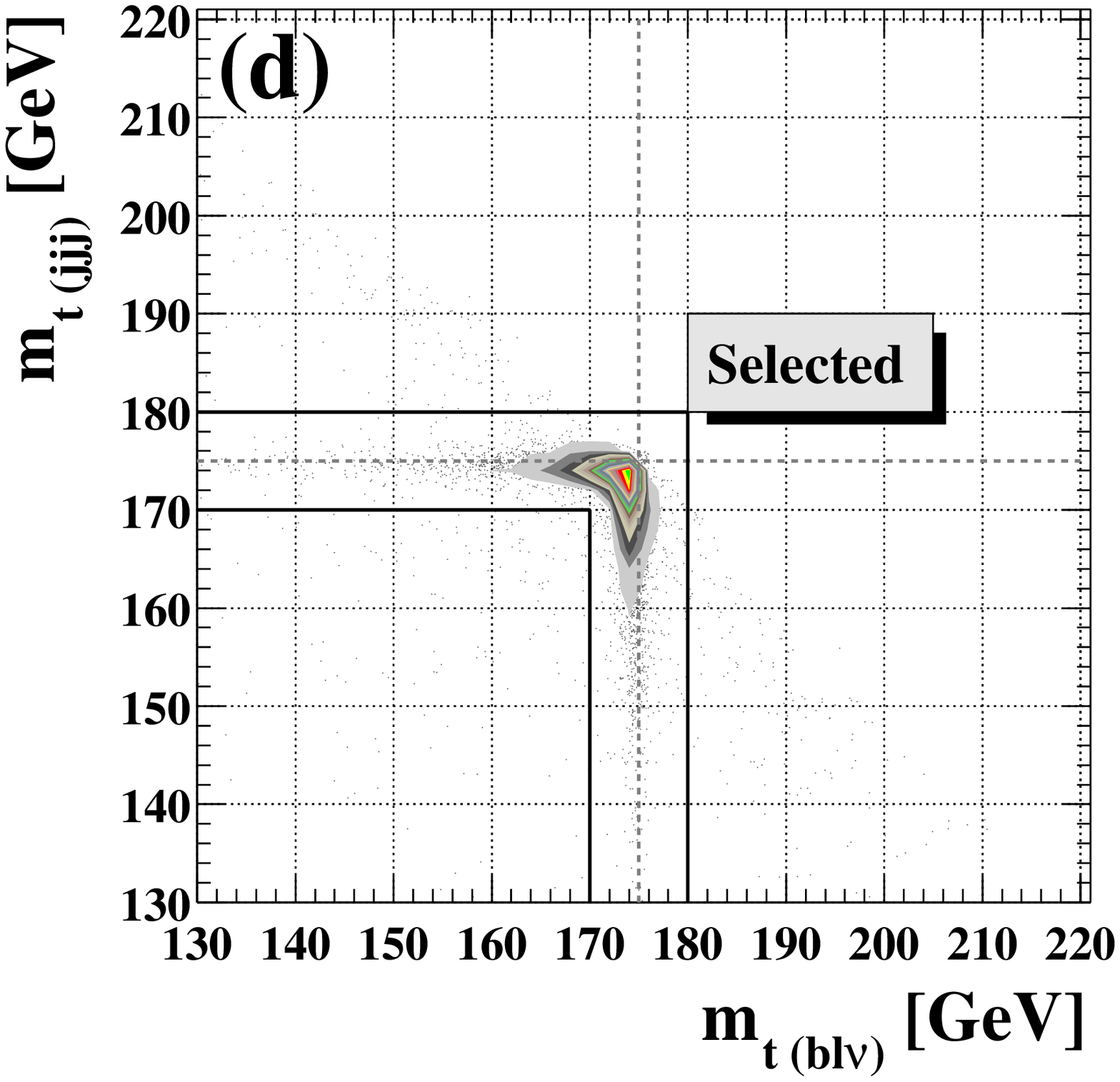}
    \includegraphics[height=5cm,clip]{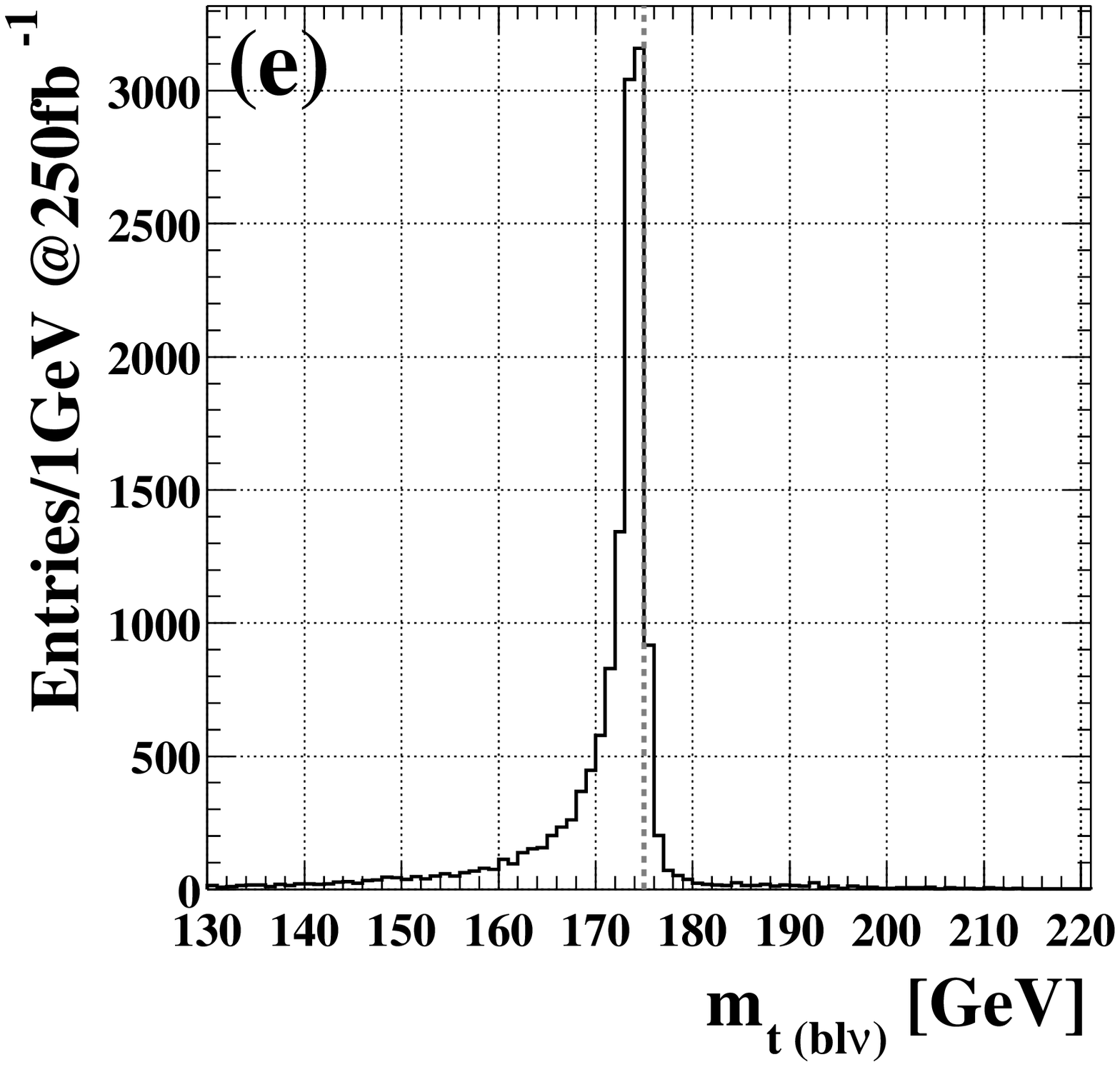}
    \includegraphics[height=5cm,clip]{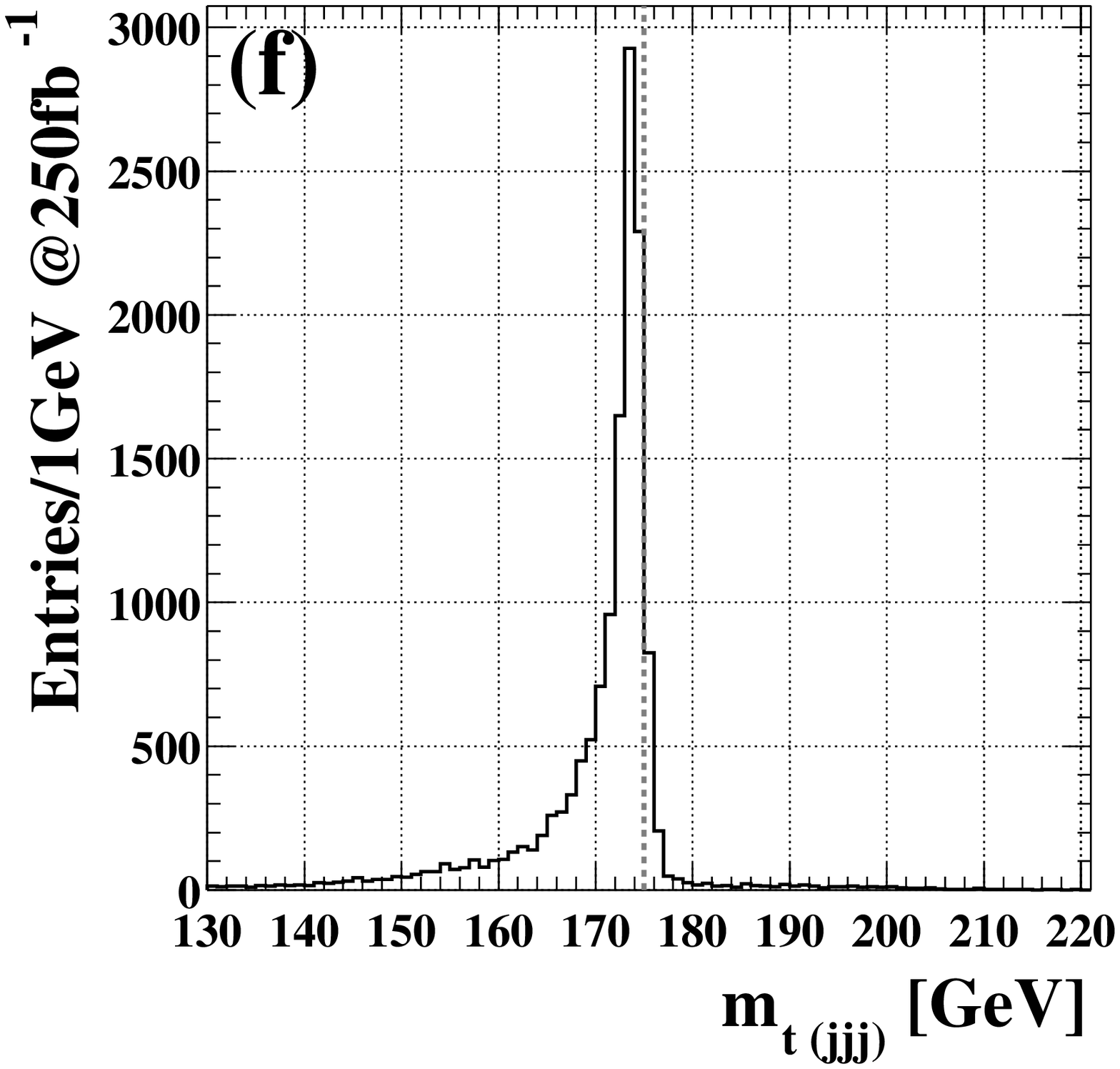}
    \\
    \includegraphics[height=5cm,clip]{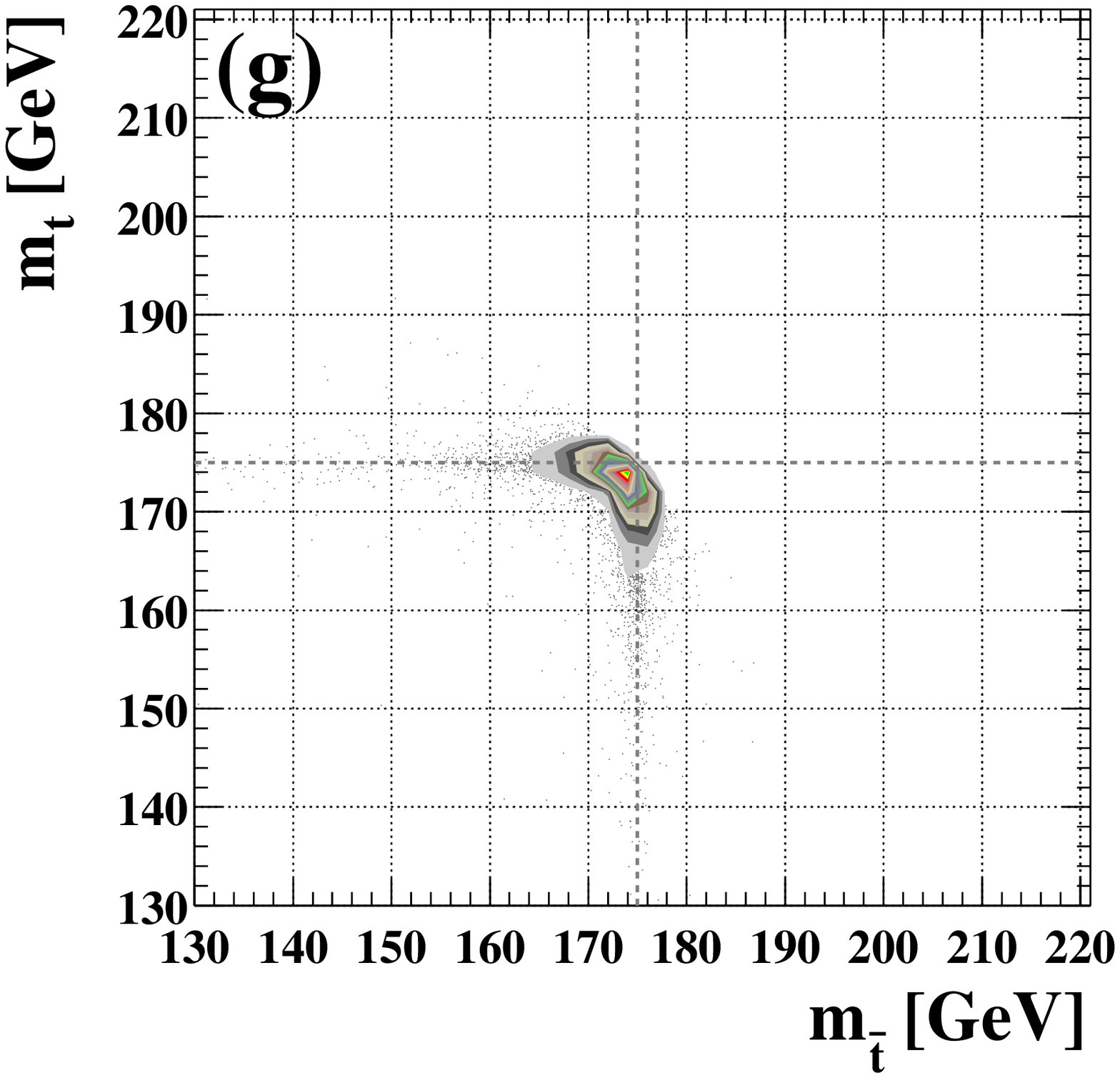}
    \includegraphics[height=5cm,clip]{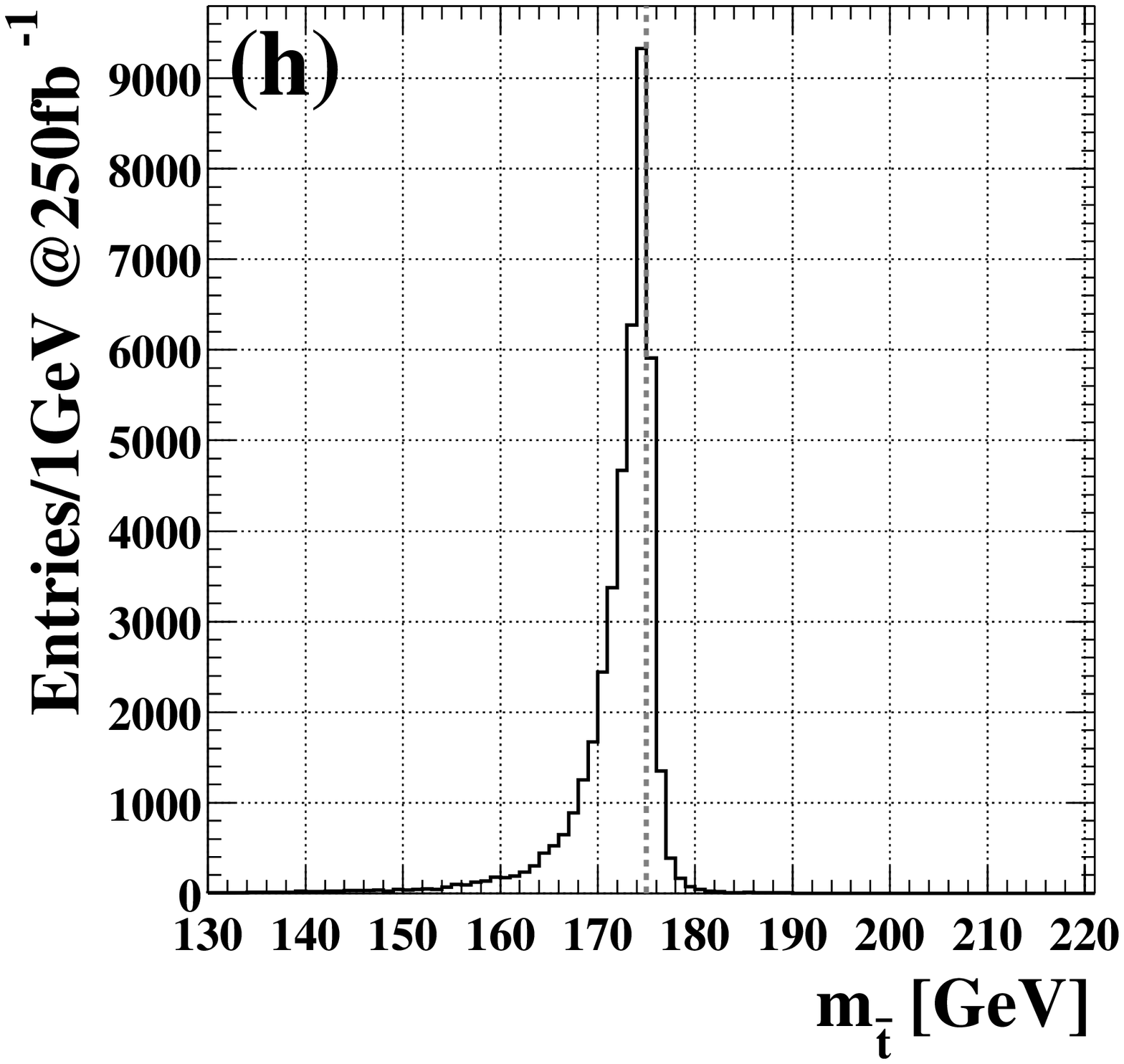}
    \includegraphics[height=5cm,clip]{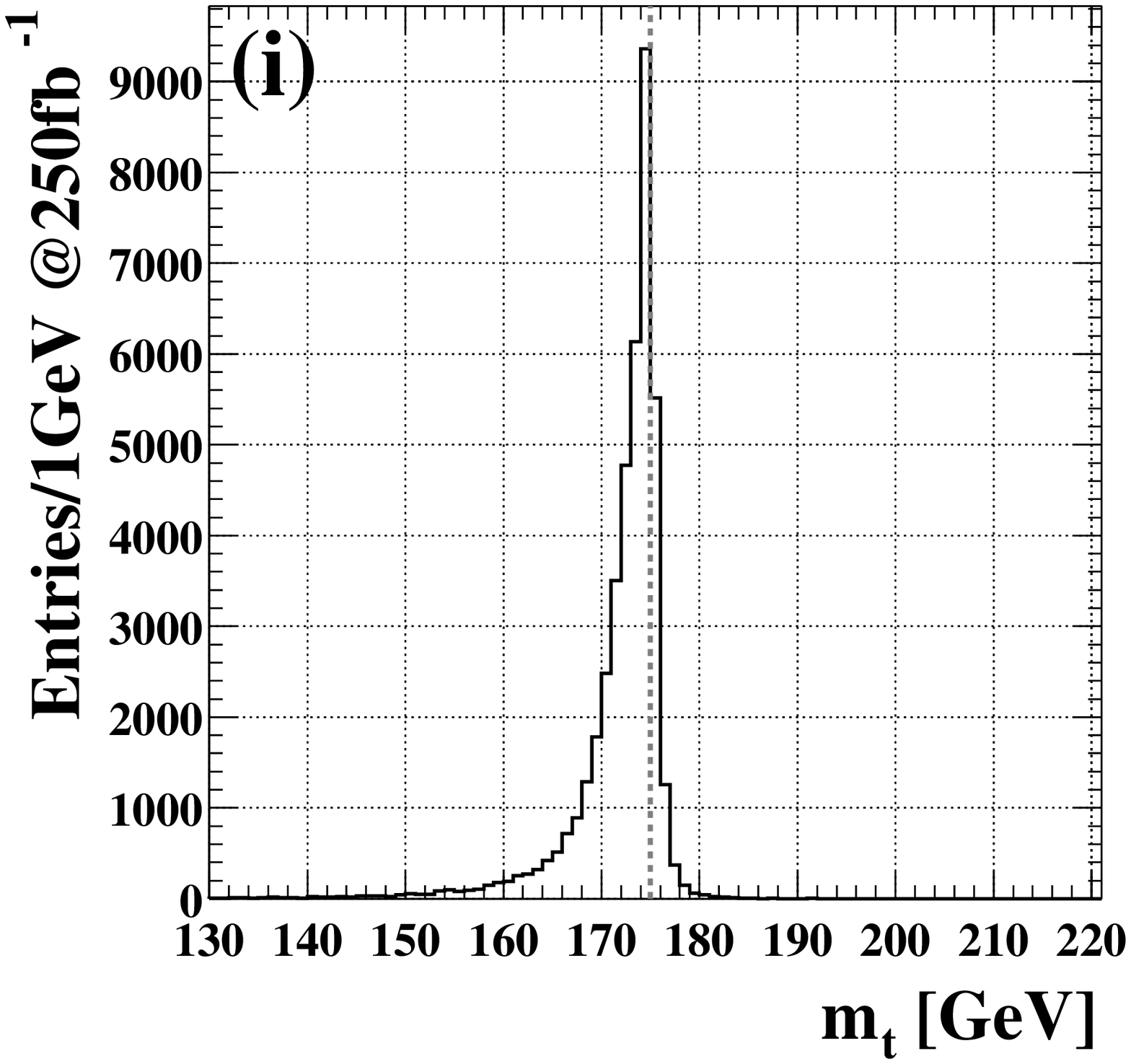}
  \end{center}
  \caption{
  	Scatter plot of the reconstructed $t$($\bar{t}$) mass
  	from 3 jets versus that from a lepton plus a $b$-jet
  	(a) before the kinematical fit,
  	together with (b) its horizontal/$b\ell\nu$
  	and (c) vertical/3-jet projections.
  	(d) through (f) are similar plots after the kinematical
  	fit and (g) though (i) are corresponding plots for
  	generated values.}
  \label{Mttbar}
\end{figure}

Fig.~\ref{Mttbar}-a) plots the reconstructed mass for
the $t$($\bar{t}$) decayed into 3 jets against
that of the $\bar{t}$($t$) decayed into a lepton plus a $b$ jet,
before the kinematical fit.
The strong negative correlation is due to the fact that
the neutrino from the leptonically-decayed $W$ is
reconstructed as the total missing momentum.
Figs.~\ref{Mttbar}-b) and -c) are the projections
of Fig.~\ref{Mttbar}-a)
to the horizontal and vertical axes, respectively,
showing systematic shifts of the peak positions.\footnote{
This is in contrast with the result in~\cite{Fujii:1993mk},
where a quite tight set of cuts
was imposed upon the reconstructed $W$ and $t$ masses,
and consequently their peak shifts were less apparent
at the cost of significant loss of usable events.
The goal of this study is to establish an analysis procedure
to restore those events which would have been lost,
by relaxing the tight cuts while keeping reasonable accuracy for
event reconstruction.
}
Figs.~\ref{Mttbar}-d) through -f) are similar plots to
Figs.~\ref{Mttbar}-a) through -c) after the
kinematical fitting,
while Figs.~\ref{Mttbar}-g) through -i) are corresponding
distributions of generated values (Monte-Carlo truth).
The kinematical fit sent most of the events
to the L-shaped region indicated in
Fig.~\ref{Mttbar}-d), as it should,
and made the distribution look like
the generated distribution shown in Fig.~\ref{Mttbar}-g).
Consequently, the peak shifts observed in
the Figs.~\ref{Mttbar}-b) and -c) have been
corrected as seen in Figs.~\ref{Mttbar}-e) and -f).
There are, however, still some small fraction of events
left along the minus $45^\circ$ line.
These events were so poorly measured that it was
impossible to restore.
The cut (angled region) indicated in Fig.~\ref{Mttbar}-d)
allowed us to remove them without introducing any
strong bias on the reconstruction of the kinematical variables.

\begin{figure}[htpb]
  \begin{center}
    \includegraphics[height=5cm,clip]{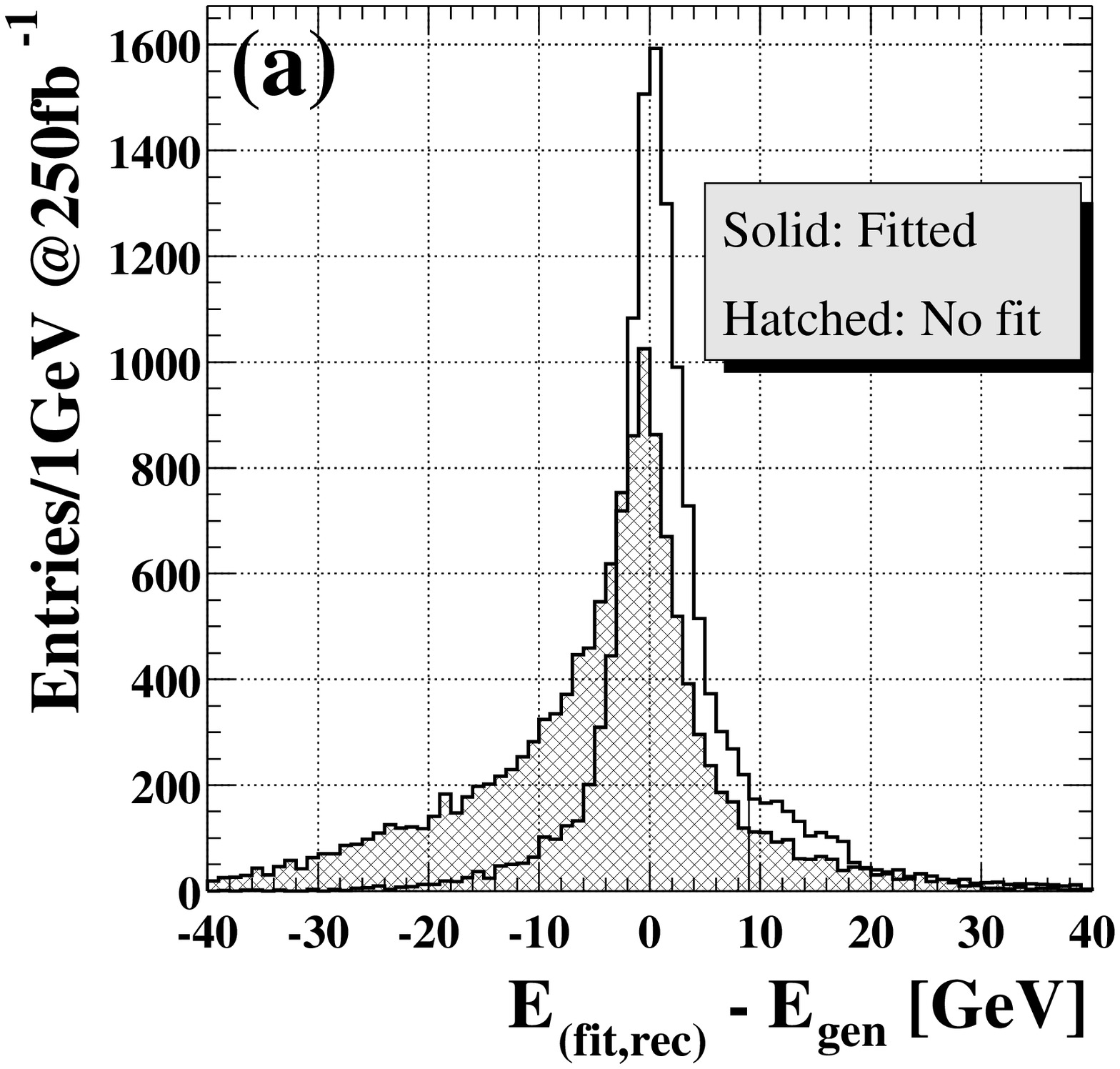}
    \includegraphics[height=5cm,clip]{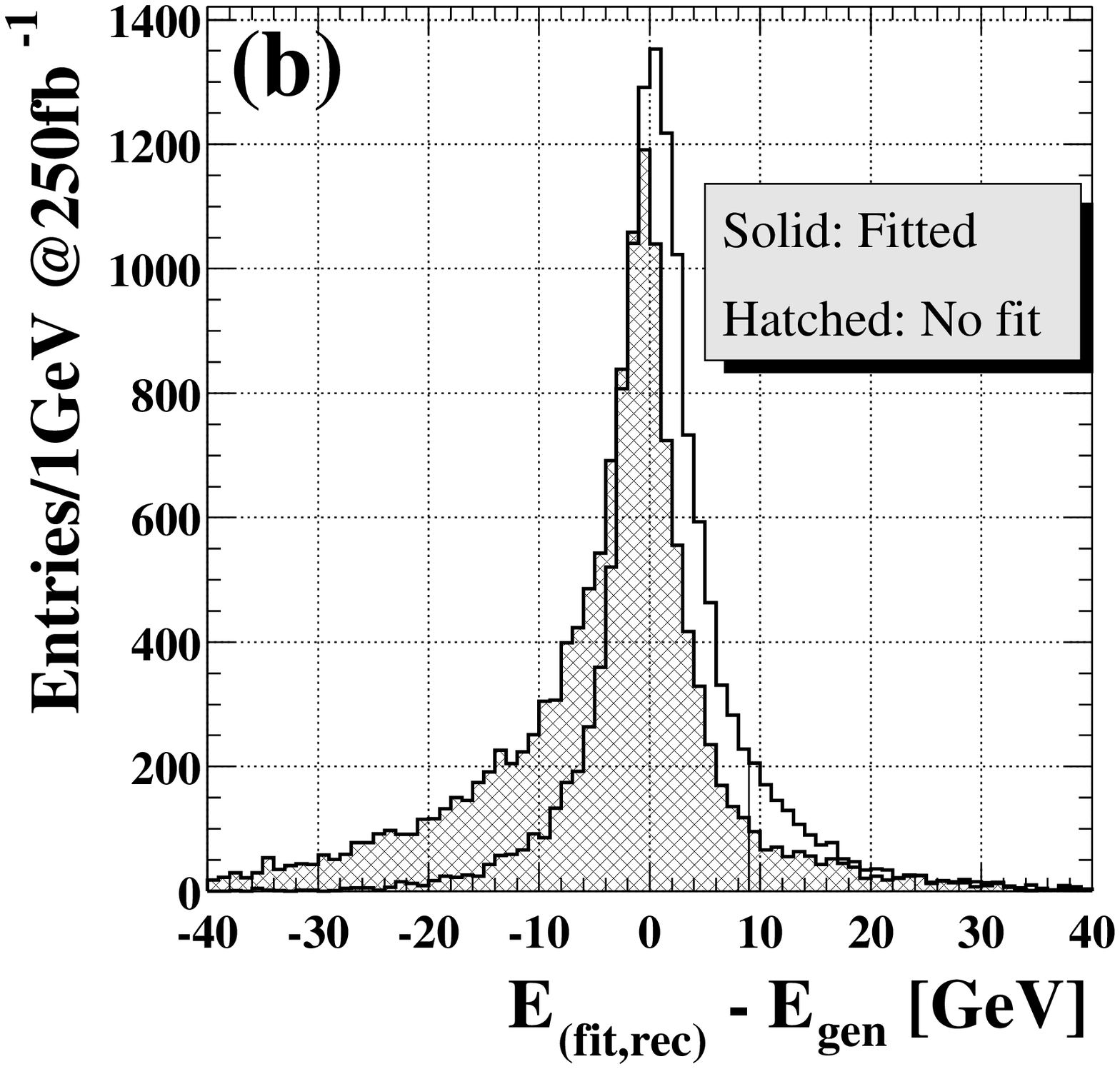}
    \\
    \includegraphics[height=5cm,clip]{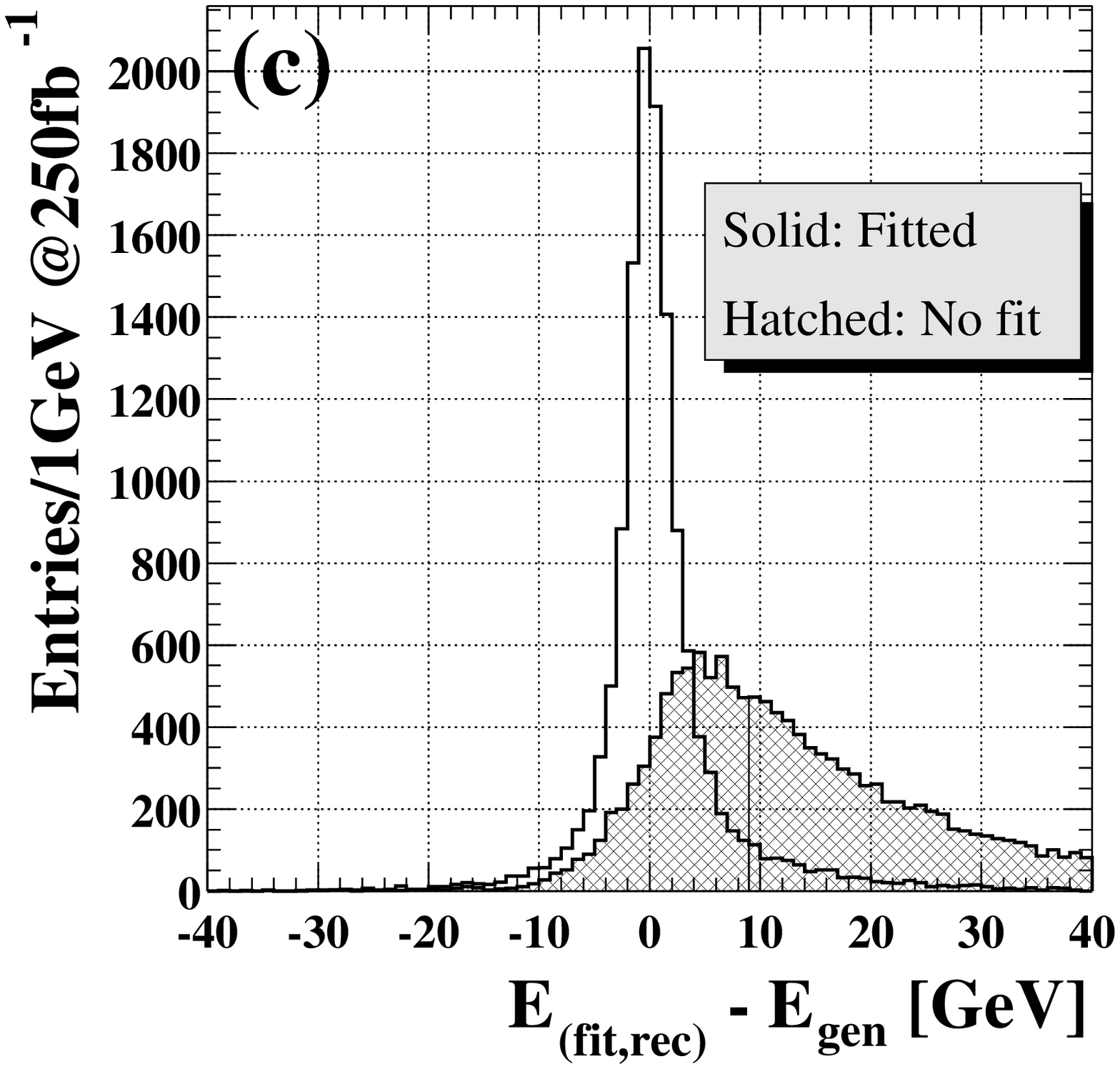}
    \includegraphics[height=5cm,clip]{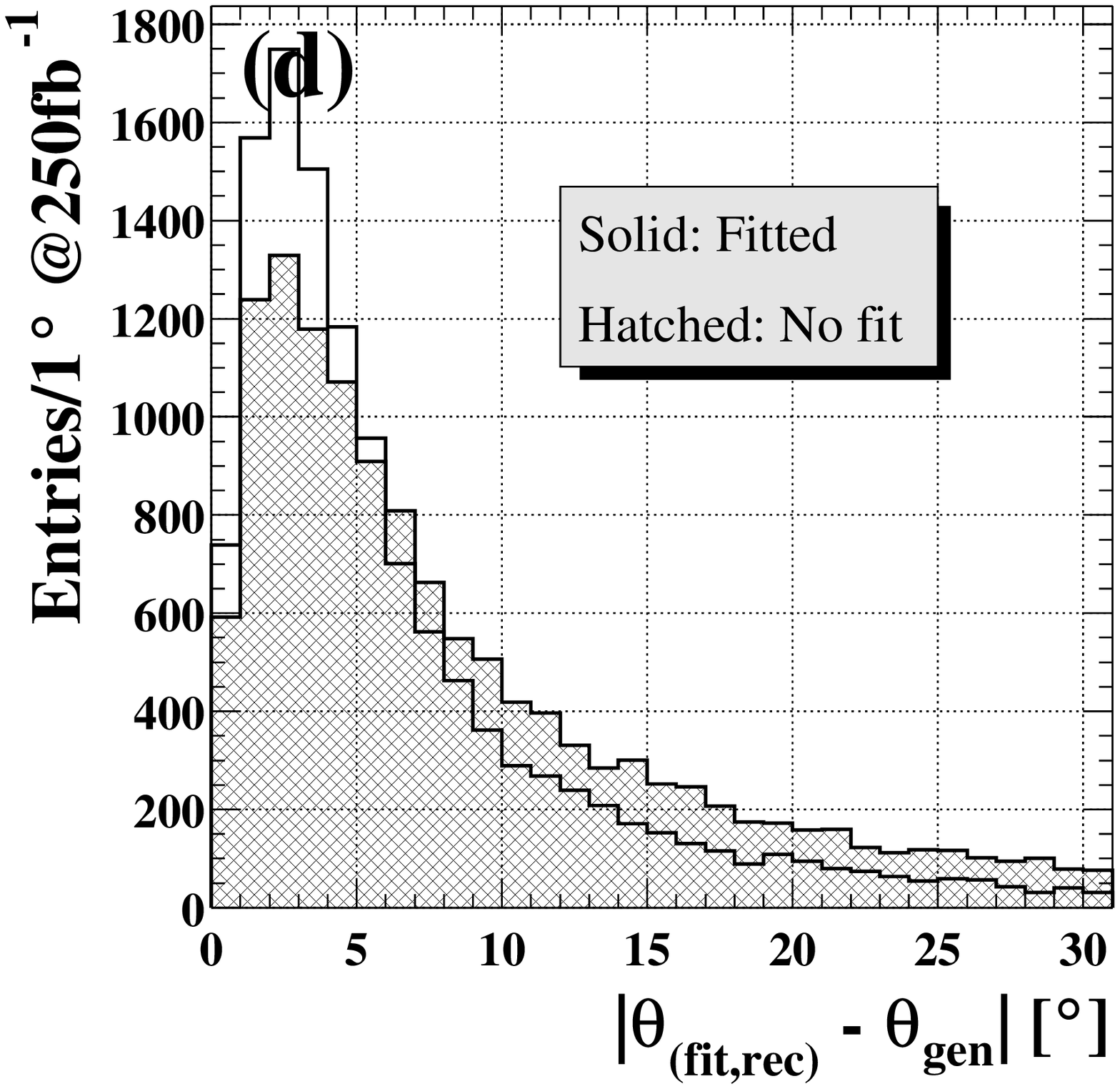}
  \end{center}
  \caption{
  	Distributions of the difference of the
  	reconstructed and generated energies
  	of the $b$ or $\bar{b}$ jet attached
  	to (a) leptonically-decayed
  	and (b) hadronically-decayed $W$ bosons,
  	together with
  	distribution of the difference of the
  	reconstructed and generated (c) energies
  	and (d) directions
  	of the direct neutrino from the
  	leptonically-decayed $W$,
  	before (hatched) and after (solid)
  	the kinematical fit.}
  \label{Eresbnu}
\end{figure}

Now the question is how the above constraints improve
the parameters of the fit such as
the energies of $b$ and $\bar{b}$ jets,
the direction and the magnitude of the
missing neutrino from the leptonically-decayed $W$,
on which we expect significant influences.
Figs.~\ref{Eresbnu}-a) and -b) plot
the difference between the reconstructed
and the generated energies of the $b$ ($\bar{b}$) quark attached
to the leptonically-decayed $W$ and
that of the $\bar{b}$ ($b$) attached to
the hadronically-decayed $W$, respectively,
before (hatched) and after (solid) the kinematical fit.
The plots demonstrate that the kinematical constraints
recover the energies carried away by neutrinos
from the $b$ or $\bar{b}$ decays.
The improvement is more dramatic for the direct neutrino
from the leptonically-decayed $W$, which is reconstructed
as the total missing momentum; see Figs.~\ref{Eresbnu}-c) and -d)
which show distributions of the difference of
the reconstructed and generated neutrino energies ($\Delta E_\nu$)
and directions ($\Delta \theta_\nu$).

\begin{figure}[htpb]
  \begin{center}
    \includegraphics[height=5cm,clip]{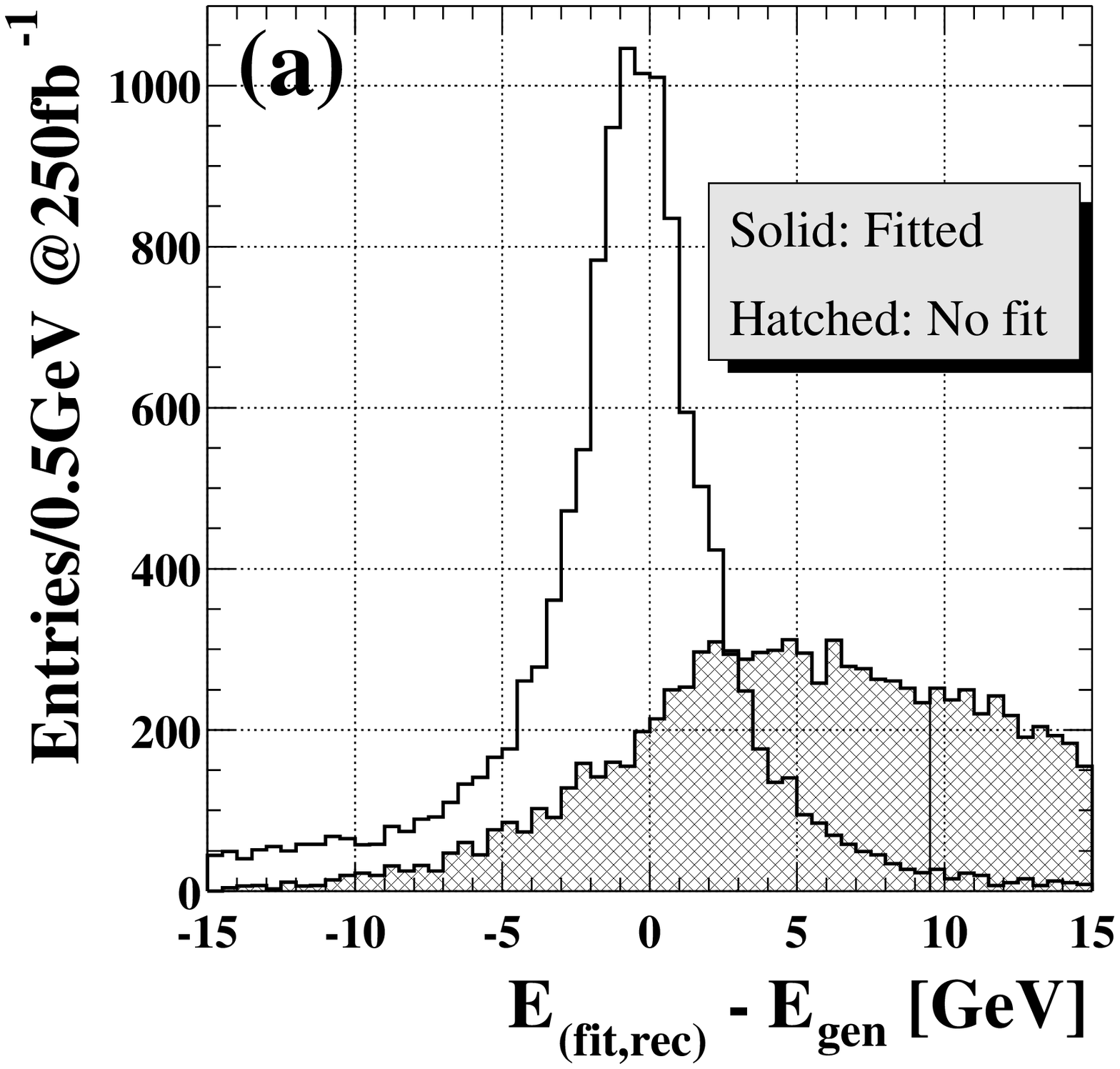}
    \includegraphics[height=5cm,clip]{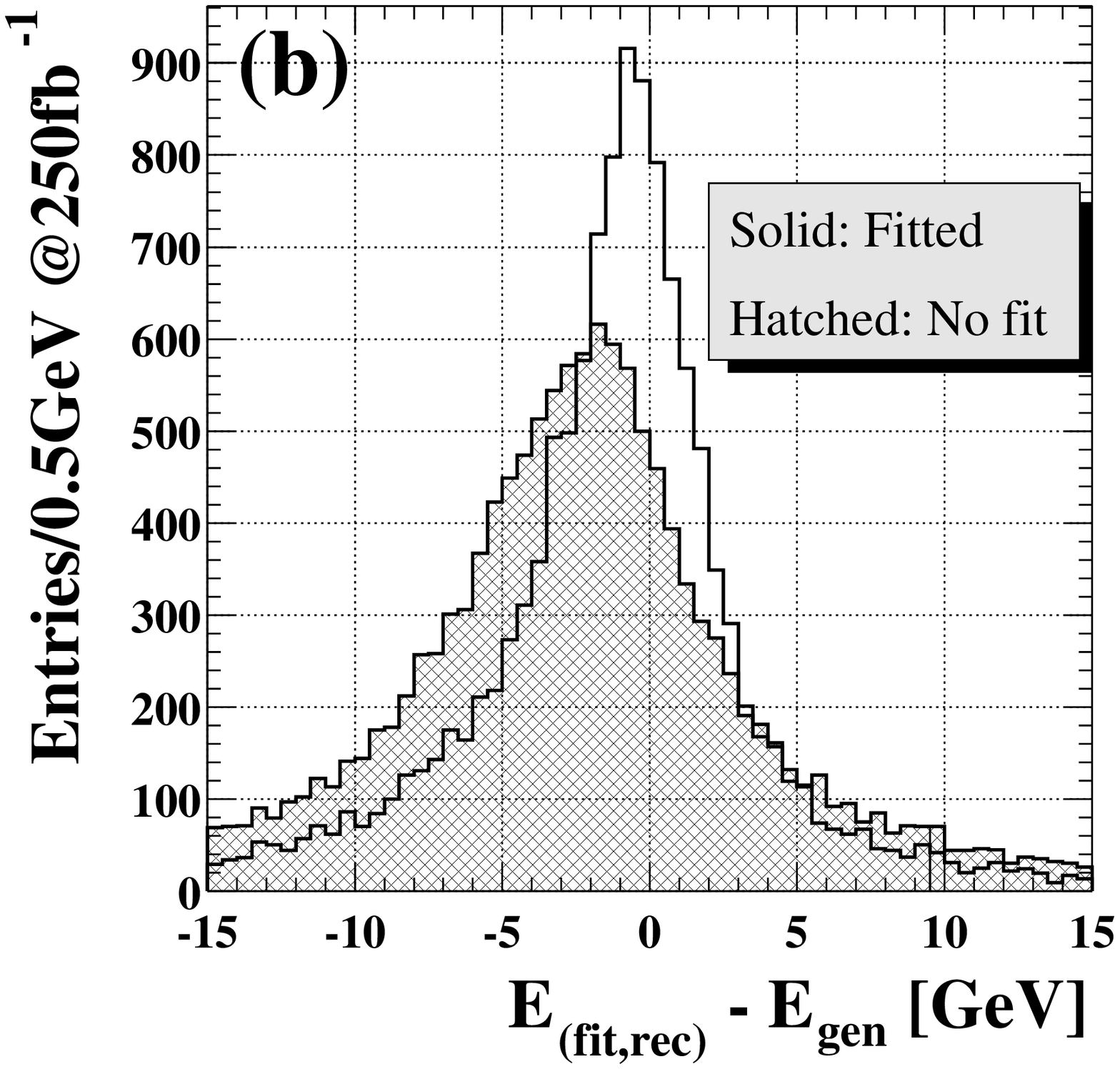}
    \includegraphics[height=5cm,clip]{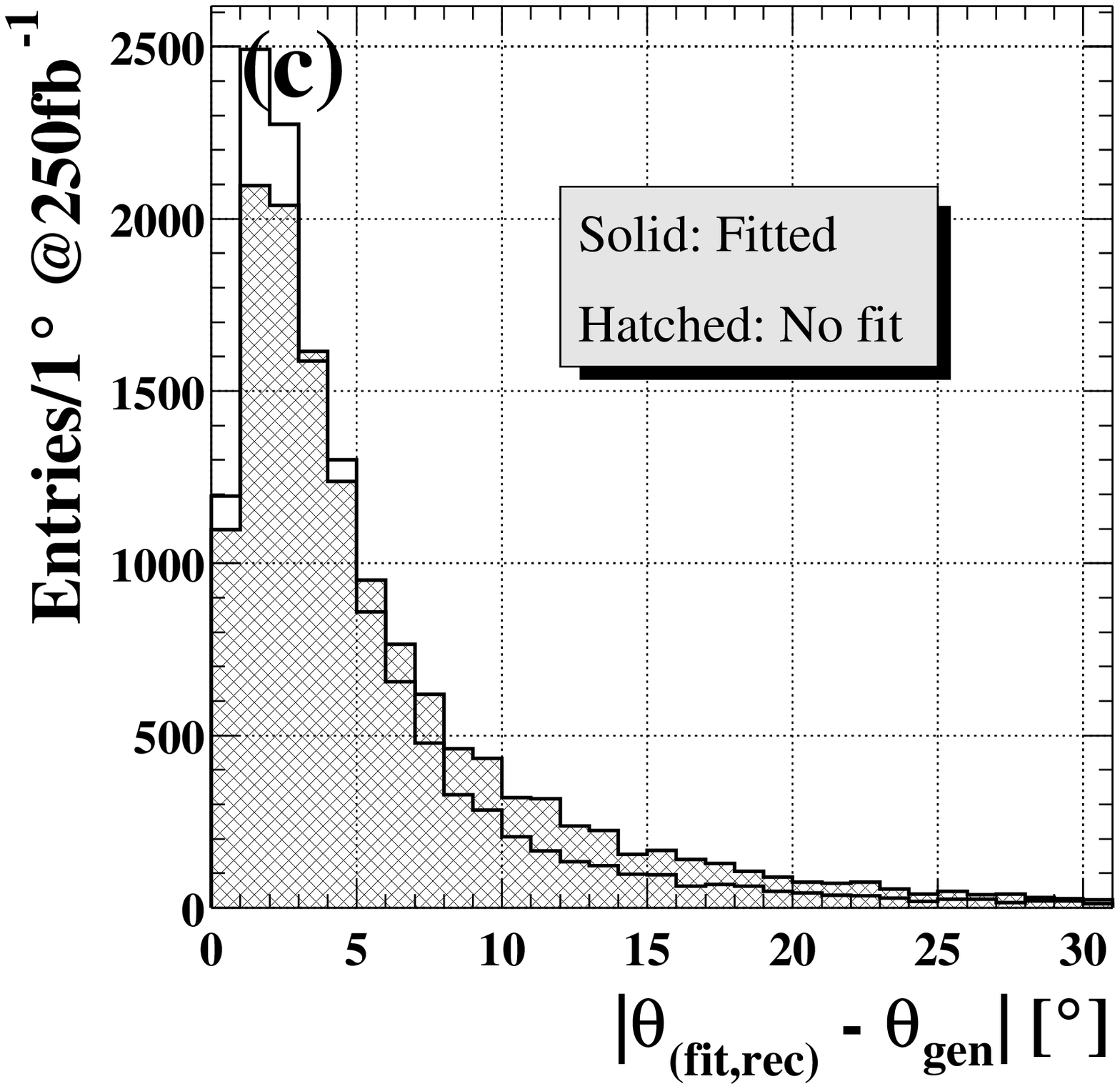}
  \end{center}
  \caption{
  	Distributions of the difference of the
  	reconstructed and generated energies
  	of (a) leptonically-decayed
  	and (b) hadronically-decayed $W$ bosons,
  	and (c) distribution of the difference of the
  	reconstructed and generated directions of
  	the leptonically-decayed $W$,
  	before (hatched) and after (solid)
  	the kinematical fit.}
  \label{EresW}
\end{figure}

The improvements in these kinematical variables
are reflected to the improvements in the
reconstructed $W$ energies and directions
as shown in Fig.~\ref{EresW}-a)
for the energy of the leptonically-decayed $W$,
-b) for the hadronically-decayed $W$,
and -c) for the direction of the leptonically
decayed $W$.
We can see dramatic improvements in all of
these distributions, although
the improvement in the direction of the hadronically-decayed
$W$ was less dramatic.

\begin{figure}[htpb]
  \begin{center}
    \includegraphics[height=5cm,clip]{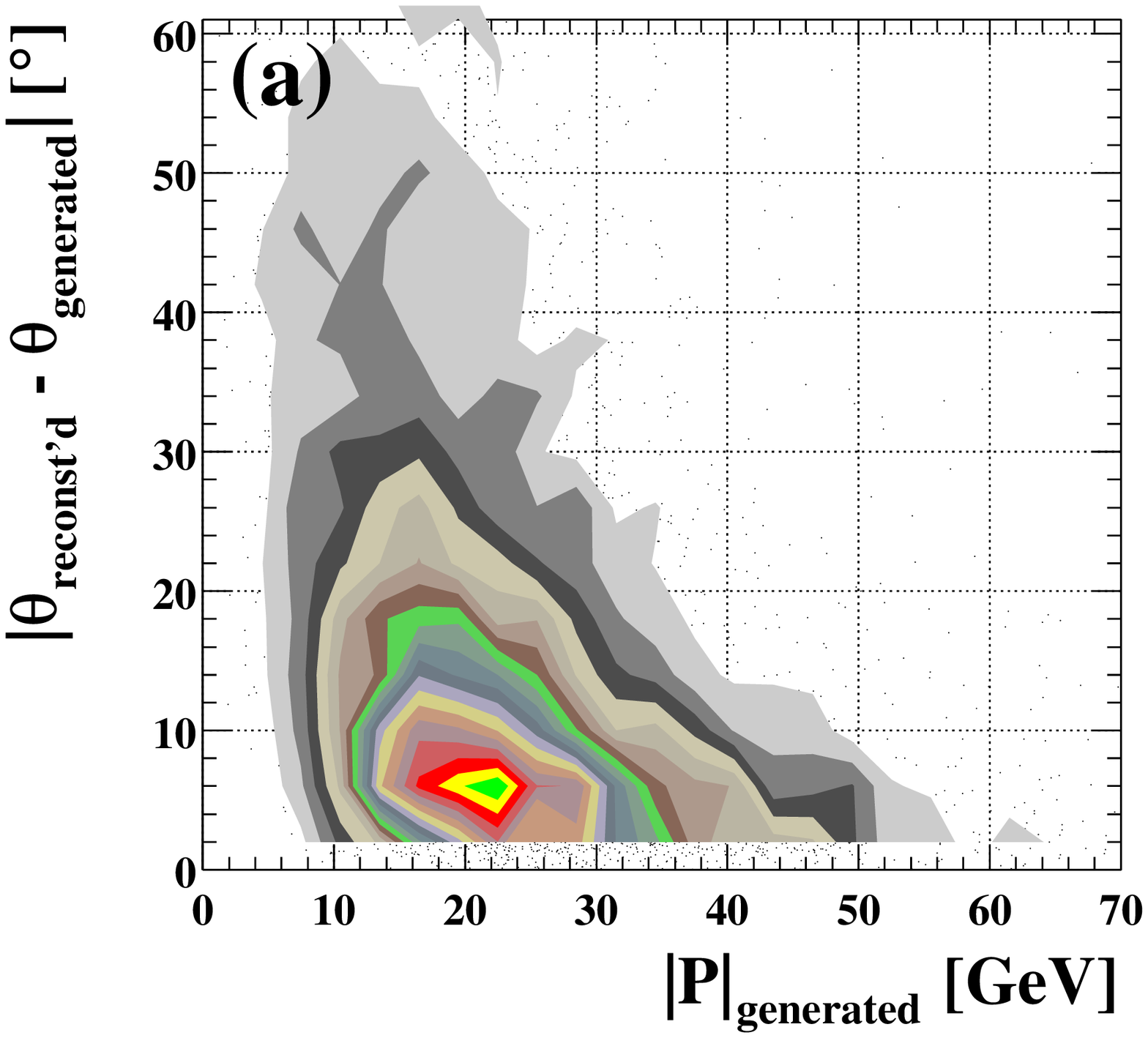}
    \includegraphics[height=5cm,clip]{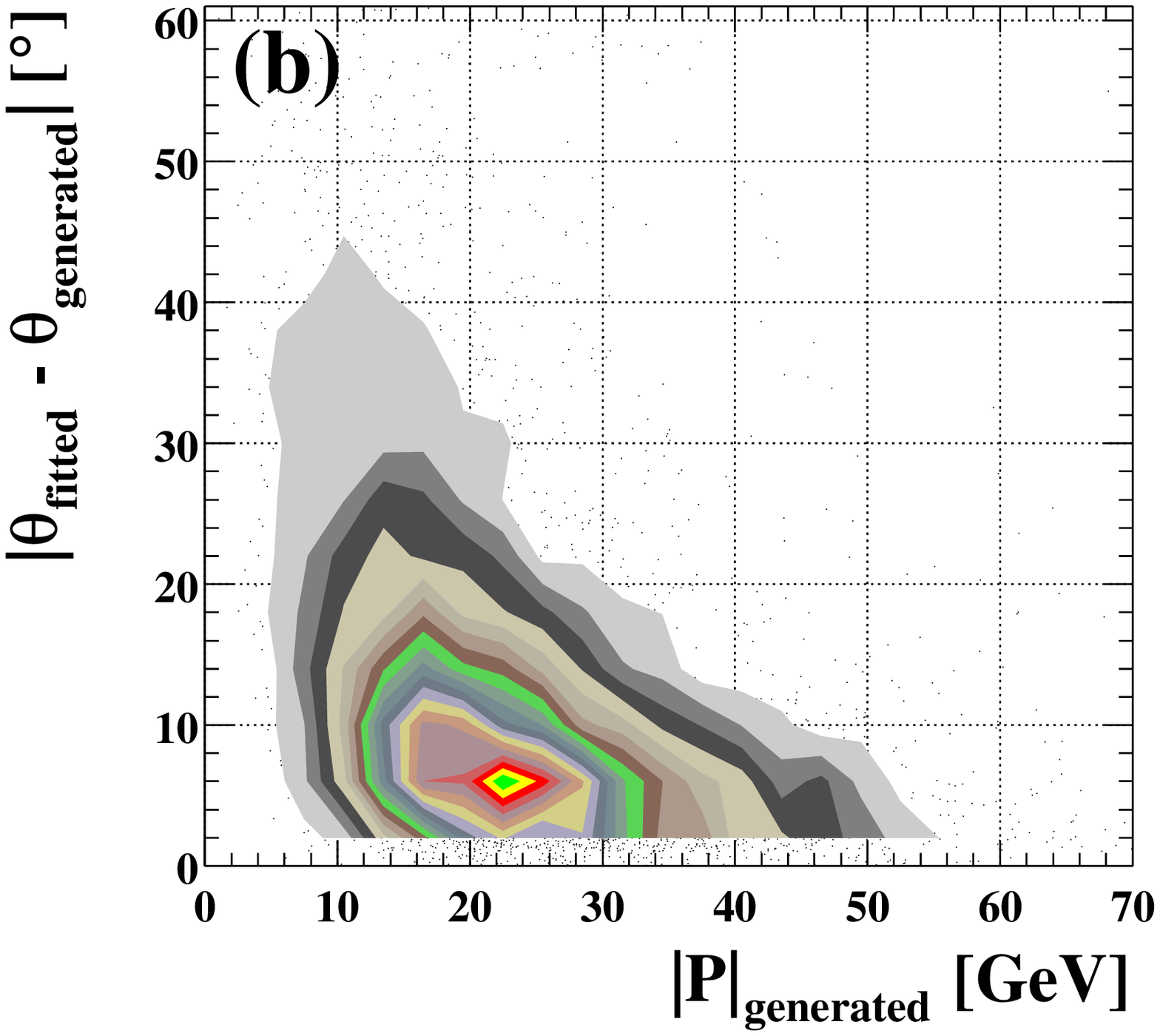}
  \end{center}
  \caption{
	The difference of the reconstructed and
	generated directions of the $t$ or $\bar{t}$ quark
	plotted against the generated top momentum,
	(a) before and (b) after the kinematical fit.}
  \label{Ares}
\end{figure}

Finally, we will examine the effects of
the kinematical fit on the measurements
of the direction and the magnitude of the
top quark momentum.
In Figs.~\ref{Ares}-a) and -b),
the difference of the reconstructed and
generated directions of the $t$ or $\bar{t}$ quark
is plotted against the generated top momentum,
before and after the kinematical fit, respectively.
We can see appreciable improvement by the fit.
Nevertheless, since the top quark direction
becomes more and more difficult to measure as the
top quark momentum decreases,
the resolution is still somewhat poor in the
low momentum region.
The angular resolution is largely determined by the
reconstruction of the $t$ or $\bar{t}$ decayed into 3 jets.
Remember that the resolution improvements
were less significant for the hadronically-decayed $W$,
since the power of the constraints was used up mostly
to recover the momentum information of the direct neutrino
from the leptonically-decayed $W$ and the energy resolution
for jets from the $W$ was left essentially unimproved.
The improvement in the measurements of the top quark direction
is mostly coming from the improvement in the $b$ or $\bar{b}$
jet measurement.
By the same token, the effect of the fit on the measurement
of the magnitude of the top quark momentum
is also less dramatic compared to that on the
leptonically-decayed $W$.

In the case of the 6-jet mode, for which
there is no direct energetic neutrino from $W$'s,
we can use the power of the constraints to
improve the jet energy measurements.
Consequently, we may expect more significant improvement
in the top quark momentum measurement.

%
\section{A Possible Application}

\begin{figure}[htpb]
  \begin{center}
    \psfrag{cosw}{\hspace{-4mm} \hbox{$\cos \theta_W$}}
    \psfrag{cosl}{$\cos \theta_\ell$}
    (a)
    \begin{minipage}{6.5cm}\centering
      \includegraphics[height=5.5cm,clip]{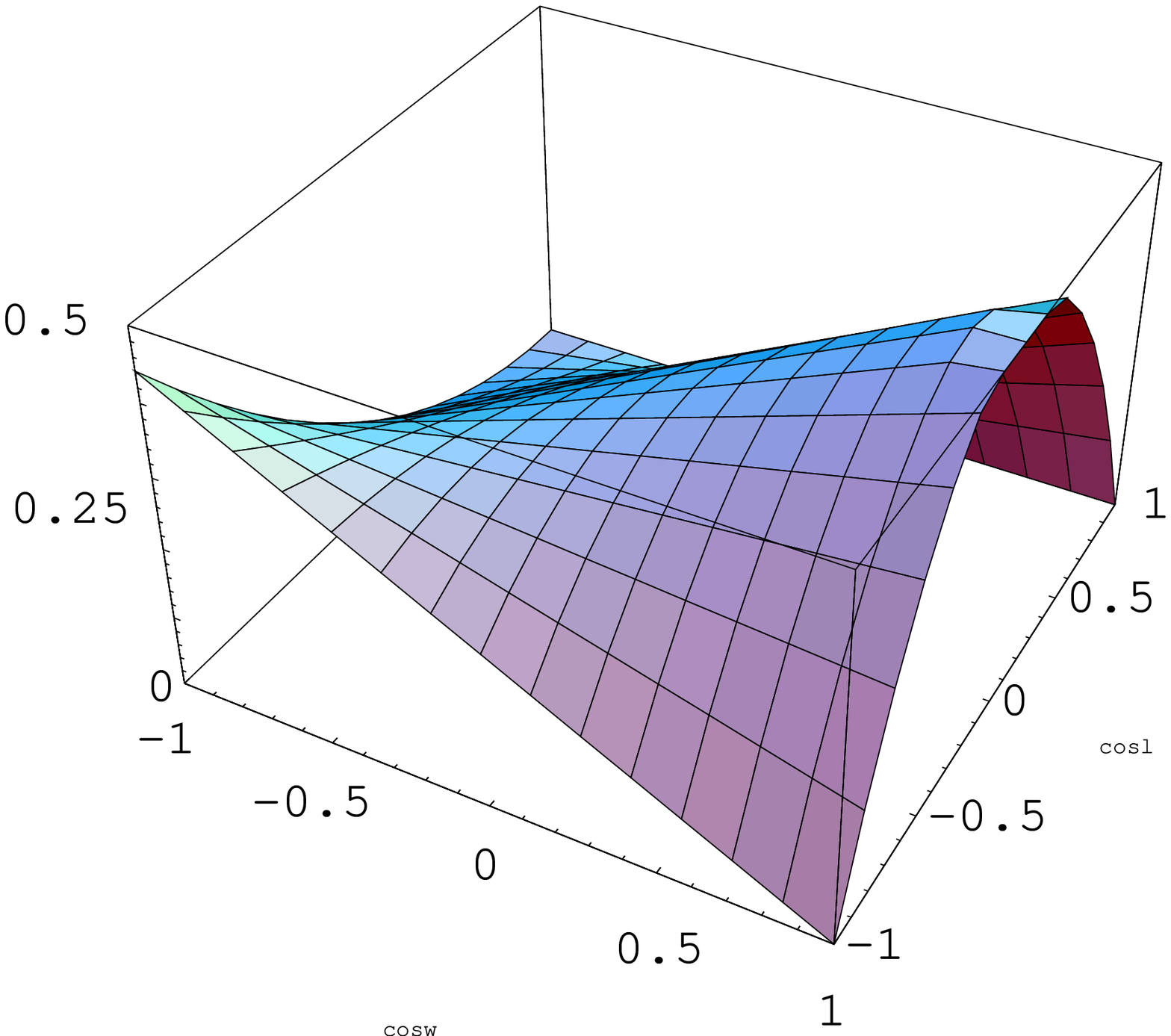}
    \end{minipage}
    \hspace*{1.0cm}
    (b)
    \begin{minipage}{6.5cm}\centering
      \includegraphics[height=5.5cm,clip]{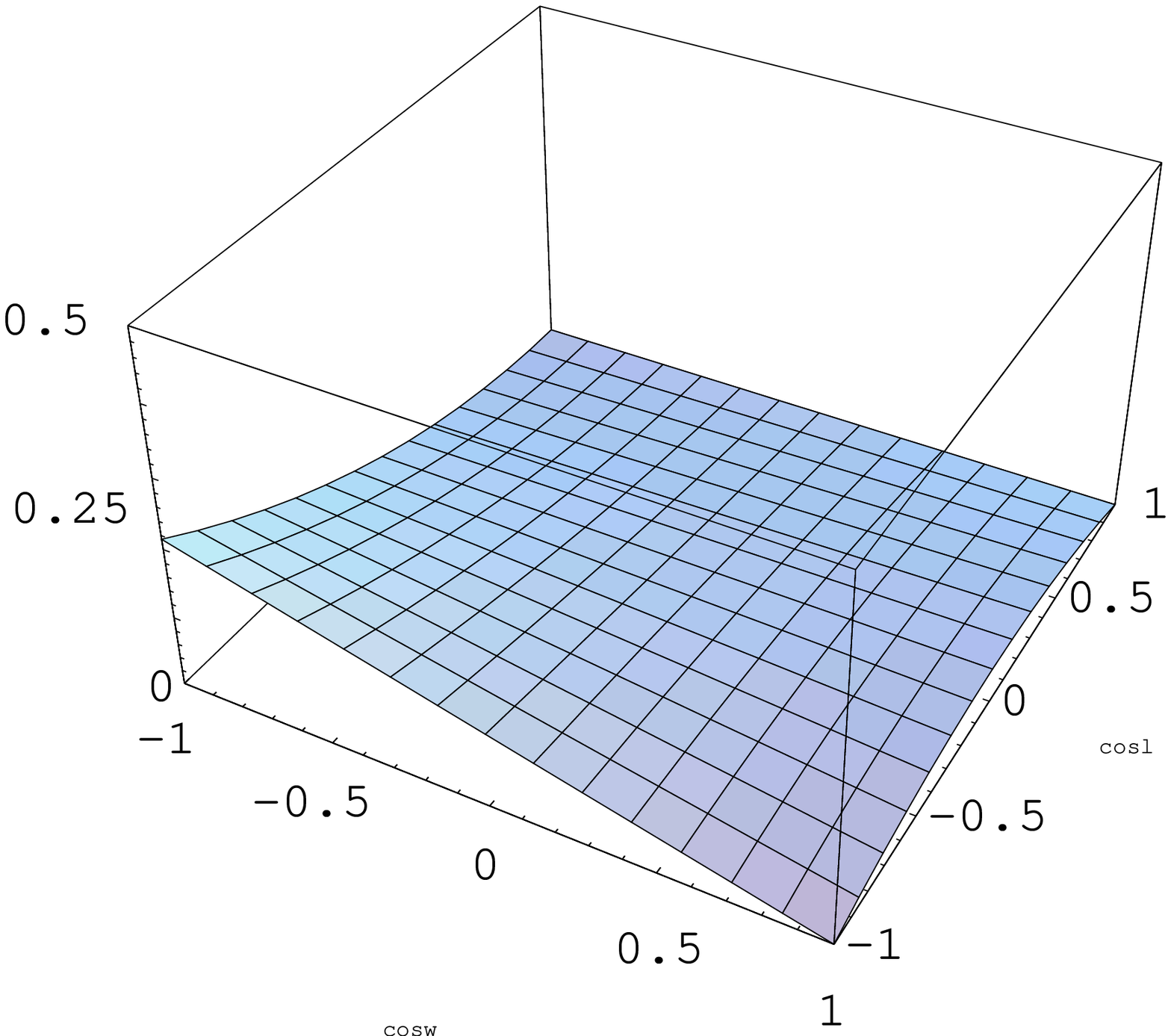}
    \end{minipage}
  \end{center}
  \caption{
(a) Normalized differential decay width
${\cal N}^{-1}\, d\Gamma(t_\uparrow \to b\ell\nu)
/(d \cos \theta_W d \cos \theta_\ell)$
for $f_2^R=0$. (b) Difference of the normalized differential decay widths
for $f_2^R=0.1$ and for $f_2^R=0$.
In both figures the differential widths are normalized by
${\cal N} = \Gamma_t \times {\rm Br}(W\to \ell \nu)$ for $f_2^R=0$.}
  \label{angular-dist}
\end{figure}

We discuss a possible application of our kinematical reconstruction
method.
Let us consider measurements of the decay form factors of the
top quark in the $t\bar{t}$ threshold region.
We assume that deviations of the top-decay form factors from the
tree-level SM values
are small and consider the deviations only up to the first order,
i.e.\ we neglect the terms quadratic in the anomalous form factors.
Then the cross sections depend only on
two form factors $f_1^L$ and $f_2^R$ in the limit $m_b \to 0$
although the most general $tbW$ coupling includes six independent
form factors~\cite{Kane:1991bg}:
\bea
\Gamma^\mu_{Wtb} = - \frac{g_W}{\sqrt{2}} \, V_{tb} \,
\bar{u}(p_b) \biggl[
\gamma^\mu \, f_1^L P_L
-
\frac{i\sigma^{\mu\nu}p_{W\nu}}{M_W}
f_2^R P_R
\biggr] u(p_t) ,
\eea
where $P_L = (1-\gamma_5)/2$ and $P_R = (1+\gamma_5)/2$.
At tree level of the SM, $f_1^L =1$ and
$f_2^R=0$.
A variation of $f_1^L$ changes only the normalization of the differential
decay width of the top quark, whereas a variation of $f_2^R$
changes both the normalization and the shape of the decay distributions.
Thus, we expect that the kinematical reconstruction is useful for
disentanglement of the two form factors and in particular for the
measurement of $f_2^R$.
For simplicity we assume $f_1^L=1$ hereafter.\footnote{
In order to determine $f_1^L$ simultaneously,
we may, for instance, use independent information from the measurement
of the top width~\cite{Fujii:1993mk}.
}
Since transverse $W$ (denoted as $W_T$)
is more sensitive to $f_2^R$ than longitudinal $W$
($W_L$), our strategy
is to extract $W_T$ using the angular distribution
of $W$ (in the rest frame of $t$) and the angular distribution
of $\ell$ (in the rest frame of $W$).
It is well known that $W_T$ is enhanced in the backward region
$\cos \theta_W \simeq -1$, where the angle $\theta_W$ is measured from the
direction of the top quark spin in the $t$ rest frame.
Also, we may enhance $W_T$ by collecting $\ell$ emitted in the
backward direction $\cos \theta_\ell \simeq -1$,
where the angle $\theta_\ell$ is measured from the
direction of $-\vec{p}_t$ in the $W$ rest frame.
These features are demonstrated in Figs.~\ref{angular-dist}:
We plot\footnote{
We used the helicity amplitudes given in \cite{Kane:1991bg} for calculating
these differential decay widths.
}
(a) the differential decay width for the decay of the top quark
with a definite spin orientation
$d\Gamma(t_\uparrow \to b\ell\nu)/(d \cos \theta_W d \cos \theta_\ell)$
for $f_2^R=0$ and (b) the difference of the differential widths
for $f_2^R=0.1$ and for $f_2^R=0$.
The plots show that we may measure $f_2^R$, for instance,
from the ratio of the numbers of events in the regions
$\cos \theta_W, \cos \theta_\ell < 0$ and
$\cos \theta_W, \cos \theta_\ell > 0$.
Since a most prominent result of our kinematical fitting is
the improvement in the measurement of
the momentum of leptonically decaying $W$,
we expect that the sensitivity to $f_2^R$ would increase after the kinematical
fitting.\footnote{
We can measure $f_2^R$ also from the energy distribution of
$\ell$.
(The angular distribution of $\ell$ is insensitive
to $f_2^R$~\cite{Grzadkowski:2001tq,Rindani:2000jg}.)
In this case, we do not need to reconstruct the $W$ momentum.
From a rough estimate,
however, we find that the method using the
$W$ and $\ell$ angular distributions has a higher sensitivity to $f_2^R$.
}

It is quite advantageous to investigate decay properties of the top quark in
the $t\bar{t}$ threshold region as compared to the
open-top region $E \gg 2 m_t$ because of several reasons.
It is useful that the top quark can be polarized close to 100\%
in the threshold region~\cite{Harlander:1994ac,Harlander:1996vg}.
This is clear in the above example.
Furthermore, we are almost in the rest frame of the top quark.
In the above example, the top quark is highly polarized in its
rest frame.
Hence, the event rate expressed in
terms of $\cos\theta_W$ and $\cos\theta_\ell$ is
a direct measure of the amplitude-squared,
$|\sum_{i=L,T} {\cal A}(t_\uparrow \to b W_i) \times
{\cal A}(W_i \to \ell\nu)|^2$ (without phase-space Jacobian),
which allows for simple physical interpretations of event shapes.
On the other hand, we do not gain resolving power
for the decay form factors by raising the c.m.\ energy.
This is in contrast with the measurements of the $t\bar{t}$ {\it production}
form factors.

%
\section{Summary and Conclusions}

To make maximum use of future $e^+e^-$ linear colliders'
experimental potential,
the top quark reconstruction in the lepton-plus-4-jet mode has been studied
under realistic experimental conditions of $e^{+}e^{-} \to t\bar{t}$ process
near its threshold.
As a new technique to fully reconstruct $t\bar{t}$ final states,
we have developed a kinematical fitting algorithm which aims to
reconstruct the momentum-vector of top quark more accurately.

The missing energy carried away by neutrinos from bottom quark decays
has been recovered by the kinematical fitting.
However, the effects of the kinematical fitting on
the top quark momentum are not as dramatic as we wanted.
This is because the top quarks are almost at rest in the threshold region
and therefore their momenta are difficult to measure.
Moreover, in the lepton-plus-4-jet mode
many constraints are used up by recovering the information on the neutrino
from leptonically-decayed $W$ bosons.
On the other hand,
the remarkable improvements of the energy resolution of $b$-jets
and the angular and energy resolutions of leptonically-decayed $W$'s
have been achieved by the kinematical fitting.
These improvements should benefit the form factor measurements in general.
As a possible application,
we considered measurements of decay form factors
including $f_2^R$, on which correct reconstruction of
the leptonically-decayed $W$ may have a large impact.

There have been a number of
theoretical studies on measurements of the top-quark production and
decay form factors using the $e^+e^- \to t\bar{t}$ process.
Many of these analyses assumed either the most optimistic case or the most
conservative case with respect to the kinematical reconstruction
of event profiles.
In the former case, one assumes that the momenta of all the particles
(including $t$ and $W$) can be determined precisely;
in the latter case, one uses only partial kinematical information,
e.g.\ the direction of $b$, and the energy and momentum of $\ell$.
Our analysis indicates that both assumptions
are not the most realistic approximations in actual experimental situations.

As further studies, we will move on to investigate expected sensitivities
to various anomalous couplings in the lepton-plus-4-jets mode systematicaly.
We also plan to extend the kinematical fitting algorithm
to the 6-jet mode, where we expect more significant improvements
of the momentum resolution of the top quark and combinatorial background.

%
\vskip 0.5cm
\begin{flushleft}
\underline{\bf{Acknowledgements}}
\end{flushleft}
\vskip 0.4cm

The authors wish to thank all the members of the ACFA working group
for useful discussions and comments.
In particular, they are grateful to S.~D.~Rindani for valuable
discussions on strategies for measurements of
top quark's possible anomalous couplings,
and A.~Miyamoto for improving JSF (JLC Study Framework)
to incorporate their requests.
This work is partially supported by JSPS-CAS
Scientific Cooperation Program under the Core University System
and the Grant-in-Aid for Scientific Research No.12740130 and No.13135219
from the Japan Society for the Promotion of Science.

%
%

\def\plb#1#2#3{{\it Phys.~Lett.~}{\bf B#1}, #2 (#3)}
\def\prd#1#2#3{{\it Phys.~Rev.~}{\bf D#1}, #2 (#3)}
\def\prl#1#2#3{{\it Phys.~Rev.~Lett.~}{\bf #1}, #2 (#3)}
\def\zpc#1#2#3{{\it Z.~Phys.~}{\bf C#1}, #2 (#3)}

\end{document}